\documentclass[preprint2]{aastex}
\usepackage{graphicx,float,amssymb,natbib}

\newcommand{\hii}{H\,{\sc ii} }

\newcommand{\siii}{Si\,{\sc ii} }
\newcommand{\siiii}{Si\,{\sc iii} }
\newcommand{\siiv}{Si\,{\sc iv} }
\newcommand{\hei}{He\,{\sc i} }
\newcommand{\heii}{He\,{\sc ii} }
\newcommand{\mgii}{Mg\,{\sc ii} }

\newcommand{\cai}{Ca\,{\sc i} }
\newcommand{\fei}{Fe\,{\sc i}}
\newcommand{\feii}{Fe\,{\sc ii} }
\newcommand{\tiii}{Ti\,{\sc ii}}
\newcommand{\oiii}{O\,{\sc iii}}
\newcommand{\oii}{O\,{\sc ii} }
\newcommand{\ciii}{C\,{\sc iii} }
\newcommand{\cii}{C\,{\sc ii} }
\newcommand{\niii}{N\,{\sc iii} }  
\newcommand{\niv}{N\,{\sc iv} }

\newcommand{\nv}{N\,{\sc v} }

\newcommand{\noprint}[1]{}

\shorttitle{Spectroscopic study of the N159/N160}
\shortauthors{Fari\~na et al.}

\begin{document}

\title{SPECTROSCOPIC STUDY OF THE N159/N160 COMPLEX IN THE LARGE MAGELLANIC CLOUD}

\author{Cecilia Fari\~na\altaffilmark{1} and Guillermo L. Bosch}
\affil{Facultad de Ciencias Astron\'omicas y Geof\'{\i}sicas, Universidad Nacional de La Plata,\\ 1900 La Plata, Argentina \\ IALP-CONICET, Argentina}

\author{ Nidia I. Morrell}
\affil{Las Campanas Observatory, Observatories of the Carnegie Institution of Washington,\\ La Serena, Chile}

\author{Rodolfo H. Barb\'a}
\affil{Complejo Astron\'omico El Leoncito, Avda. Espa\~na 1412 Sur, San Juan, Argentina\\ Departamento de F\'{\i}sica, Universidad de La Serena, Benavente 980, La Serena, Chile}

\and

\author{Nolan R. Walborn}
\affil{Space Telescope Science Institute, 3700 San Martin Drive,
Baltimore, MD 21218, USA}

\altaffiltext{1}{ceciliaf@fcaglp.unlp.edu.ar}

\begin{abstract}
We present a spectroscopic study of the N159/N160 massive-star forming
region south of 30~Doradus in the Large Magellanic Cloud, classifying a
total of 189 stars in the field of the complex. Most of them belong to O
and early B spectral classes; we have also found some uncommon and very
interesting spectra, including members of the Onfp class, a Be P\,Cygni
star, and some possible multiple systems.
Using spectral types as broad indicators of evolutionary stages, we considered the evolutionary status of the region as a whole.
We infer that massive stars at different evolutionary stages are present throughout the region, favoring the idea of a common time for the origin of recent star formation in the N159/N160 complex as a whole, while sequential star formation at different rates is probably present in several subregions.
\end{abstract}

\keywords{Magellanic Clouds --- HII regions --- individual:(N159/N160) --- galaxies: starburst --- stars: classification, early-type}

\section{INTRODUCTION}
The N159/N160 complex comprises a star-forming region located in the Large Magellanic Cloud (LMC), approximately 600~pc in projection south of 30 Doradus. The complex extension, from north to south is approximately 15\arcmin\,(225~pc) \citep{2005AJ....129..776N}. Several reasons motivate the present study of the N159/N160 complex. The most important to mention are:

\begin{enumerate}
\item The whole complex is populated by young massive stars and it presents
  numerous features characteristic of active star formation.
\item  Its distance (approximately 50 kpc for the LMC) is small enough to
  permit the study of individual objects (small scale), but also large enough to
  allow a broad view of the distribution of stars and interstellar matter over
  the whole complex (large scale).
\item There has been little spectral classification study of stars within the
  complex. Aside from a few conspicuous objects, spectral classes of
  individual stars have been inferred only from photometry or indirectly from
  UV photon fluxes.
\end{enumerate}

The whole complex has been widely observed and studied at different frequencies. The \hii regions in N159/N160 were first catalogued by \cite{1956ApJS....2..315H}. \cite{1981MNRAS.197P..17G} discovered the first extragalactic protostar in this complex, and the same year, \cite{1981MNRAS.194P..33C} found the first Type I extragalactic OH maser.
\cite{2001PASJ...53..985Y} performed a global study of CO in the LMC, focused on giant molecular clouds and their association with young objects, and they found a compact group of the youngest clusters ($t\leq$10 Myr) located near N159 while those with ages between 10 and 30 Myr extend to the northern part of the complex (between 30 Dor and N159).
Candidate Herbig Ae/Be and OB stars were detected in N159/N160 by \cite{2005AJ....129..776N}, who found indications of sequential cluster formation within the complex and, on a major scale, from N160 to N159S. These authors also suggest a scenario for sequential formation.
There are numerous studies on subregions within the complex. Most of them
focused on N159 or part of it, as did, for example \cite{1992A&AS...94..359D},
\cite{1992A&A...259..480D}, who obtained {\it UBVRI}, $H\alpha$, $H\beta$, and [\oiii] photometry of N159A.
Additional {\it JHK} stellar photometry is available from \cite{2004A&A...422..129M} and \cite{1982A&A...111L..11H} observed at $H\alpha$, $H\beta$, and [\oiii]; {\it UBVI} stellar photometry of the whole region can be found in the Magellanic Clouds Photometric Survey by \cite{2004AJ....128.1606Z}.
\cite{2000ApJ...545..234B} studied the carbon in the  gaseous
phase of the whole complex; the authors described  N159/N160 as composed of
three distinct and well separated regions, aligned from north to south,
 N160 lies in the northern part, N159 is at the center, and N159S is in the southern part, and they support the idea that different stages of stellar formation are being observed in these regions, which would be currently active in N159, while not yet present in N159S.
However, the stellar component of the region has been the subject of independent studies using various techniques and covering different subregions. This circumstance might be biasing the global interpretation of what is taking place in this vast star-forming region. We have therefore conducted spectroscopic observations and performed a preliminary analysis of stars in the N159/N160 complex as a contribution to a better understanding of the star forming history of the region.\\
Section~2 contains a description of the observations, data processing and reductions. In Section~3 we present the spectral classification as well as comments on some individual interesting spectra. Section~4 presents a brief discussion and interpretation of the data.
An Appendix summarizes the most important classification criteria and the meaning of special characters used in some spectral types.

\section{OBSERVATIONS, DATA PROCESSING AND REDUCTIONS}
\subsection{Spectroscopic Observations}
The images were taken during two consecutive nights in 2003 November, at
the Las Campanas Observatory (LCO), Chile, with the 2.5\,m Ir\'en\'ee du
Pont telescope and the Wide Field Reimaging CCD Camera, in the
multi-object spectroscopic mode. The detector was a Tek$\#5$ CCD with
2048x2048 (24~$\mu$m) pixels providing images of 25~arcmin diameter at a
scale of 32.3 arcsec\,mm$^1$ (0.774 arcsec\,pixel$^{-1}$). The H\&K grism,
centered at 3700~\AA, was used resulting in spectrograms with reciprocal dispersion of 1.3\,\AA\,px$^{-1}$ and $\sim$ 3.1\,\AA\ resolution.
We used nine different slit masks grouping stars of similar brightness, selected in order to produce the minimum overlapping among the resulting spectra.
Each mask has an average of 25 slits of variable length and 1 arcsec wide.
The number of spectra obtained in each mask depends on the number of stars falling in each slit.
Individual exposures were 1200\,s long; two or three exposures were taken for each slit mask to combine into the final spectra.

\subsection{Source Selection}
No optical photometry was available at the time the multislit
masks were made, therefore program stars were selected from colors
calculated from their Two Micron All Sky Survey (2MASS) $JHK$ photometry. We derived an IR
analogue to the $Q$ parameter; $Q_{\mathrm{IR}}=(J-H)-(H-K)
E_{J-H}/E_{H-K}$. Although we were aware that infrared photometry is even
less sensitive than {\it UBV} for early-type stars, this parameter has behaved well in detecting OB stars in the area. Astrometry was derived using a set of stars common to our acquisition images and the 2MASS catalog.

\subsection[]{Data Processing and Reductions}
The data were reduced using the IRAF\footnote{IRAF is developed and distributed by NOAO, operated by AURA, Inc., under agreement with the NSF.} software package at the La Plata Observatory.
The bias level was removed but the images were not divided out by
flat fields because the quartz lamp exposures were obtained with the
multislit masks in and adjacent spectra slightly overlap each other.
The subtraction of background nebular emission was not trivial in some of the spectra. Two situations are worth mentioning to illustrate this point:
\begin{enumerate}
\item In our multi-object spectroscopy masks several stars
did not fall in the center of the
    slit, but rather near one of the ends, making it difficult to take a
    background sample large enough to be properly removed. This results
    in background contamination that is not easy to remove due to the fact that
    the nebular background emission in which the program stars are
    immersed is bright and highly variable on small spatial scales. Many
    of these are stars that were not originally included in the program
    but showed up as bonus spectra in program star slits.
\item Because the spectra are slightly curved along the dispersion direction, some spectra from consecutive slits are overlapped. This contamination from other slits prevented us from obtaining a proper sample of adjacent background emission.
\end{enumerate}
These are the reasons that in some of the spectra presented the background nebular emission is noticeable, although in most cases it does not affect the spectral classification.
The signal-to-noise ratio (S/N) covers a range from 20 for the noisier spectra to 210 for those of best quality. The histogram distribution among these values is nearly flat.

\section[]{RESULTS}
\subsection []{Spectral Classification}
The spectra were classified mainly following the criteria of the Digital Atlas of Optical Spectral Classification of OB Stars \citep{1990PASP..102..379W}. Other sources were consulted for some particular spectra, among them are the Digital Atlas of Peculiar Spectra \citep{2000PASP..112...50W}, the Spectral Classification System for the Earliest O Stars \citep{2002AJ....123.2754W} and R. O. Gray's Digital Atlas of Spectral Classification \footnote{http://nedwww.ipac.caltech.edu/level5/Gray/frames.html}. In Figure Set~1, the best quality spectra from our data set are presented, ordered by sequence of spectral type and luminosity class.
In Table \ref{tab:ourspec}, all the stars for which we obtained a
spectral classification are listed, along with their J2000 celestial
coordinates and {\it V} magnitudes from \cite{2004AJ....128.1606Z}.
Correlation between our star list and the photometric catalog was made
by coordinates and we exclude magnitudes for three stars that seem to be
misidentified: star 1 is very close to the limit of our field image and
stars 41 and 42 are very close together, immersed in a bright background.
In Table \ref{tab:speclit}, we have listed our program stars that had already
been classified by other authors. As mentioned in the introduction, only a few
of the stars in our sample have previous spectral classifications based on
spectroscopic observations, although many of them are included in several
studies, see \cite{2004A&A...422..129M, 1993AJ....106.1005D, 2002A&A...381..941H, 2005ApJ...620..731J} among others. As can be seen, only nine stars were included in \cite{1970CoTol..89.....S} and the spectral class given is no more precise than ``OB'' type, except for star 7. A few stars were also found in common with other catalogs of spectral classification through careful review of finding charts, since the difference in accuracy between the 2MASS coordinates and those in old catalogs made it impossible to search objects by coordinates.


\subsection{Comments on Individual Interesting Spectra}

Star~33 (Figure~8) has a peculiar spectrum, which was classified as O9.7~Iabpe. It presents emission at H$\gamma$ and all the lines appear peculiarly weak for a supergiant star.

Star~35 (Figure~9) is a widely studied object because it is the optical counterpart to the X-ray source LMC X-1. \cite{1983ApJ...275L..43H} found it to be an O7 star with a binary period of approximately 4 days; they also suggested that the secondary could be a black hole. It was later classified as O7-9~III by \cite{1984IAUS..108..317P} and \cite{1985A&A...146..242B}. We find a type of O8(f)p, where the peculiarity refers to the reversed \heii$\lambda$4686 emission. For this reason it has been associated with the Onfp class by Walborn et al. (2009, in preparation), of which the \heii profile is a defining characteristic, although in this object the absorption lines are not broadened, and the \heii profile may be related to the X-ray binary system.

Star~37 (Figure~4) was classified as O8 here, although \cite{1991ApJ...373..100C}
classified it as O3-O6, based on the presence of \heii $\lambda\lambda$4200,
4541, 4686 absorption lines and possible \niii $\lambda\lambda$4634-40-42
emission lines. The stellar \hei absorption lines were obliterated by the
very strong nebular emission lines in their observation. For our
classification we used mainly the ratios of \heii
$\lambda$4541/\hei $\lambda$4387 and \heii $\lambda$4200/ \hei
$\lambda$4144. We also used the oversubtracted nebular [\oiii] $\lambda$4363
line present in the final spectrum to estimate the nebular oversubtraction
(see \cite{1999A&AS..137...21B}). We found a correction of less than 15\%,
which indicates that the stellar \hei $\lambda$4471 absorption is indeed
deeper than \heii $\lambda$4541.  In addition, the \ciii $\lambda$4070
absorption blend is present. All of these considerations indicate
that the spectral type of star~37 must be closer to O8. Unfortunately, our
wavelength coverage for this object did not allow us to check for the presence
of \niii emission, as suspected by \cite{1991ApJ...373..100C}. The luminosity
class could not be assessed due to the lack of the \heii $\lambda$4686 line.

Star~38 (Figure~17) is a B3 supergiant that at one time was also a candidate for identification with LMC X-1. Fortunately, the weak \heii$\lambda$4686 emission seen in high S/N digital data was not visible in the early observations, or this misidentification would have been considered definite! \cite{2006ApJ...641..241R} have clearly shown that this anomalous emission feature actually arises in an unusual high-ionisation nebulosity in the field, which may be related to the X-ray source.

Star~39 (Figure~8) was classified as O9.7~Ia+pe. It is certainly an unusual spectrum, probably related to the Ofpe/WN9 or WNL class. See for comparison the spectrum of HD~152408 \citep{2000PASP..112.1243W}. However, these characteristics have not been seen before at such a late O type.

Star~41 (Figure~1) is the hottest star we have found among the sample. It was classified as O3~Vz((f*)), although the z nature (\heii$\lambda$4686 absorption stronger than other He lines, possibly an indicator of subluminosity and extreme youth) is marginal. \niii$\lambda\lambda$4634-40-42 and \niv$\lambda$4058 are clearly detected in emission and \nv$\lambda\lambda$4604-20 absorption lines are also present. \siiv$\lambda\lambda$4089-4116 seem to be in emission but the features are not conspicuous.

Star~93 (Figure~10) was classified as BN0\,Iabp due to enhanced \niii and deficient C and O lines, characteristic of OBN stars \citep{1971ApJ...164L..67W, 1976ApJ...205..419W}. The additional peculiarity is the presence of P\,Cygni profiles at H$\gamma$ and \hei lines.

Star~179 (Fig.~10) was classified as B\,Iae\,P\,Cyg. The P\,Cygni features can be seen in H$\epsilon$ and H$\delta$, but not in H$\gamma$ which is totally in emission. This object is quite similar to HD~87643 \citep{2000PASP..112...50W}, which belongs to the group of Iron Stars, of which $\eta$~Carinae is the most famous member.

Stars~54, 70, 76, 107, and 112 (Figure~20) show either anomalously broad or asymmetric profiles in their absorption lines, suggesting a possible binary nature. The wavelength resolution does not allow us to reliably analyse radial velocity information. Higher resolution spectra would be needed to assess the possibility of these being SB2 stars.

Stars~72, 82, and 151 (Figure~9) belong to the Onfp category \citep{1973AJ.....78.1067W}. In their spectra, the broad emission in \heii$\lambda$4686 with absorption reversal can be observed (see Appendix). Stars~72 and 82 are also denoted with ``+'' due to \siiv$\lambda$4116 emission; they also present C\,{\sc iii} $\lambda$4650 in emission. A luminosity class cannot be inferred for these stars because of the peculiar \heii$\lambda$4686 profiles (Walborn et al. 2009, in preparation).

\subsection{Discussion}
30~Doradus is the current optical ``hotspot'' near the northeastern edge
of a much larger, oval region several kpc in diameter, itself northeast
of the LMC Bar, in which star formation has been active for at least
10$^8$~yr. This region contains half the WR content of the LMC,
numerous young clusters of various ages, many red supergiants, and
SN~1987A.  The field south of 30~Dor is particularly rich in clusters, associations, and nebulae, including the N159/N160 complex studied here. This field also contains the most massive CO concentrations in the LMC, so it may be expected that in a few million years 30~Dor's successor will appear there.
In order to analyse the current, global evolutionary trends within the N159/N160 complex,
the observed stars have been grouped following qualitative
criteria based on those defined and used in \cite{1997ApJS..112..457W}; see also \cite{1999A&AS..137...21B}. We have defined three groups of spectral types and luminosity classes. These groups comprise distinct age ranges in massive stellar evolution. Then from the total of 189 spectra in the field of N159/N160, we have selected those that belong to one of the following three groups.
\begin{itemize}
\item Young phase: O3-5~V-I and O~Vz.
\item Middle-age phase: O6-9~V-I and B0~I.
\item Evolved phase: B0-2~V-III and B1-8~I.
\end{itemize}
Of course, some of the main-sequence stars in the second and third groups could actually belong to the younger ones, but this grouping is a first approximation for our purposes. The three groups are plotted in a direct image of the field in Figure \ref{fig:distribution}, \ref{fig:distributionN159}, \ref{fig:distributionN160},and \ref{fig:distributionN160E}, with different symbols for each. If a global sequential star formation process were taking place (as suggested by CO studies mentioned in Section~1 and also studies of neighboring regions like N158 \citep{1998A&AS..130..527T}), we would expect to find systematic trends in the object distribution according to their evolutionary stages. Indeed, there is evidence for such effects in Figure \ref{fig:distribution}: there is a clear tendency for the two younger phases to be associated with the nebulae and clusters, while the majority
of the third phase are in the dispersed northeast quadrant of the field, which is devoid of nebulosity. On the other hand, young phase objects are found in the northern part of the complex (N160)
as well as in the southern part (N159), while some more evolved objects also
appear in the latter.
Of course, we are looking at a two-dimensional projection of three-dimensional structures. It is quite likely that the younger concentrations are embedded in and/or projected against a field of more evolved and dispersed objects, just as occurs in 30~Dor itself \citep{1997ApJS..112..457W, 1999A&A...347..532S}. Also, some isolated very young objects may be runaways from the young clusters. Our data favor a scenario in which star formation was initiated throughout the region, perhaps in the event producing the kpc-scale structure noted above, and that this strong burst led to subsequent, localized star-formation events wherever sufficient remanent molecular material was available, in an ongoing process that continues at the present time. This scenario is in good agreement with other studies of subregions in the N159/N160 complex, as in \cite{2002A&A...381..941H} where the presence
of compact high-excitation \hii blobs demonstrates that this is indeed a very
young subregion likely containing different (early) evolutionary stages. Stellar spectra alone do not provide the definitive ages that are required to
construct a quantitative star formation history of the region. A complete
photometric study will be essential to disentangle the overlapping populations
and their extinctions. Such data are available from \cite{2004AJ....128.1606Z} and will be extended by new observations we are currently undertaking.
Subsequently, we shall undertake a more detailed analysis including H-R diagrams and comparison with the extensive H$\alpha$, IR, and CO observations available for this region.\\



We acknowledge comments by referee Dr. Margaret Hanson which improved the presentation of this paper. This research has made use of  Aladin \citep{2000A&AS..143...33B} and the {\sc SIMBAD} database, operated at CDS, Strasbourg, France. For this publication we had made use of NASA's Astrophysics Data System as well as data products from the Two Micron All Sky Survey, which is a joint project of the University of Massachusetts and the Infrared Processing and Analysis Center/California Institute of Technology, funded by the National Aeronautics and Space Administration and the National Science Foundation. Financial support for the authors was provided by PIP 5697-CONICET. R.\,B.\,acknowledges partial support from Universidad de La Serena, Project DIULS CD08102. N.\,W.\,acknowledges support from the STScI Director's Discretionary Research Fund. 

\newpage

\appendix
\section[]{CLASSIFICATION CRITERIA FOR OB STARS}

{\it General remarks.} For spectral types from O3 to B0\,V, the principal horizontal
(spectral type or temperature) classification criteria
are \heii$\lambda$4541/\hei$\lambda$4471 and \mbox{\heii$\lambda$4200/\hei(+\heii) $\lambda$4026}. At the later types \heii$\lambda$4541/\hei$\lambda$4387 and \mbox{\heii$\lambda$4200/\hei$\lambda$4144} are useful checks, since the former ratios become very small (but the latter ratios are also sensitive to luminosity).\\
For early B-type spectra, the principal horizontal classification criterion
shifts from the He ionization ratio to those of Si, first \siiii$\lambda$4552/\siiv$\lambda$4089 and then \mbox{\siii$\lambda\lambda$4128-4130/\siiii$\lambda$ 4552}.\\
In O-type spectra (earlier than O9), luminosity class V is defined by \heii$\lambda$4686 strongly in absorption, class III by weakened absorption in that line, class Ib by an absent or neutralized line (filled in by emission), and Ia (or just I at the earliest types) by this line in emission above the continuum. Note that there are differences of detail in this sequence as a function of the spectral type; the two-dimensional types are ultimately defined by the standards.  The designation Vz corresponds to spectra with \heii$\lambda$4686 absorption stronger than any other He lines, possibly indicating subluminosity and extreme youth (on or near the Zero-Age Main Sequence).\\
{\it Concerning the different flavors of ``f'' effects.} Of supergiants: strong \heii$\lambda$4686 and \mbox{\niii$\lambda\lambda$4634,4640-42} emission lines (Walborn 1971b). f+ denotes spectra in which \siiv$\lambda\lambda$4089,4116 are present in emission (sometimes only the latter is visible due to cancellation of absorption and emission in the former).
This category does not include objects in which the
intense lines are of P\,Cygni type (Walborn \& Fitzpatrick 2000). f* means that \niv$\lambda$4058 emission is stronger than \mbox{\niii$\lambda$4640}, and is characteristic of O2-O3 spectra \citep{1990PASP..102..379W,2002AJ....123.2754W}. Onfp Stars are characterized by broad, centrally reversed  \heii$\lambda$4686 emission, together with broadened absorption lines. There is a range in profiles from strong absorption with very weak emission wings to strong emission ``split'' by a weak absorption feature. In some cases the two emission wings are unequal in intensity. These effects may be related to rapid rotation and/or a disk structure. Only a few of these objects are known and most of them are found in the Magellanic Clouds (\cite{1973AJ.....78.1067W}; Walborn et al. 2009, in preparation). In main-sequence stars (luminosity class V), ((f)) is used to denote strong
\heii$\lambda$4686 absorption accompanied by weak \niii$\lambda\lambda$4634,4640-42 emission.
In the intermediate luminosity classes, (f) is used
when the \heii$\lambda$4686 absorption weakens and may become neutralised by emission,
while the \niii emission strength increases. Finally, Of supergiants have
both \heii and \niii strongly in emission \citep{1990PASP..102..379W}.\\
{\it Additional notes.} The ``n'' parameter describes the degree of broadening of the spectral
lines. The symbol ((n)) indicates that the lines of \siiv$\lambda$4116 and
\hei$\lambda$4121 are just merged (at $\sim1$~\AA\ resolution), and (n) represents an
intermediate case between that and the broadening in the n spectra. High rotational velocities are associated with this behavior \citep{1971ApJS...23..257W}. A ``:'' after the spectral type or luminosity class denotes, as usual, that an accurate classification cannot be given due to one of several reasons, e.g., incomplete wavelength coverage, low S/N, uncertain continuum normalization.\\
In addition to the [\oiii] and H\,{\sc i} lines, several He and Ne nebular lines are present
in the observed spectra. Among these, the \hei$\lambda$4471 line is
particularly important, as the corresponding stellar absorption line constitutes one of the
principal horizontal classification criteria for O stars (\heii$\lambda$4541/\hei$\lambda$4471
ratio). Therefore, in some cases in which the nebular background emission could
not be totally removed from the stellar spectra, a less accurate classification was achieved.\\

\newpage

\begin{deluxetable}{lccll}
\tablecaption{Spectral classification of 189 stars}
\tablehead{
\colhead{ID} & \colhead{$\alpha$ (J2000)} & \colhead{$\delta$ (J2000)} & \colhead{Spectral Type} & \colhead{$V$}
}
\startdata
1    &  5:37:40.96    &    $-$69:48:32.9    &   O9      &  \nodata\\
2    &  5:37:44.94    &    $-$69:41:59.1    &   A0 V        &  14.1\\
3    &  5:37:57.45    &    $-$69:38:16.0    &   A5 V    &  19.1\\
4    &  5:37:59.60    &    $-$69:41:13.0    &   B2.5 V    &  17.2\\
5    &  5:38:14.05    &    $-$69:38:18.1    &   B1$-$3 V    &  19.2\\
6    &  5:38:20.28    &    $-$69:44:40.6    &   F0 Ib    &  16.4\\
7    &  5:38:26.73    &    $-$69:45:52.3    &   A0 Ib    &  10.9\\
8    &  5:38:32.13    &    $-$69:37:10.4    &   G5 V    &  15.6\\
9    &  5:38:32.26    &    $-$69:35:41.5    &   B0$-$2 V    &  18.3\\
10    &  5:38:41.65    &    $-$69:35:44.1    &  A3 III   &  15.6\\
11    &  5:38:42.97    &    $-$69:35:42.8    &   A5 III    &  16.2\\
12    &  5:38:50.23    &    $-$69:46:47.9    &   A3 III$-$II    &  14.4\\
13    &  5:38:55.01    &    $-$69:38:47.4    &   A0 III$-$Ib    &  14.2\\   
14    &  5:38:57.31    &    $-$69:47:57.8    &   G0$-$3 V    &  15.7\\
15    &  5:39:02.25    &    $-$69:39:54.8    &   G0 V    &  13.3\\   
16    &  5:39:08.92    &    $-$69:42:52.4    &   B1 Vn[e]    &  15.0\\
17    &  5:39:11.86    &    $-$69:42:20.0    &   G0 V    &  15.0\\
18    &  5:39:17.72    &    $-$69:42:42.1    &   G0$-$3 V    &  16.6\\
19    &  5:39:18.93    &    $-$69:36:58.2    &   B0.7 Ib    &  14.2\\
20    &  5:39:19.68    &    $-$69:40:13.2    &   B1$-$2 V    &  17.3\\
21    &  5:39:25.74    &    $-$69:42:42.7    &   A0 V    &  17.7\\
22    &  5:39:30.42    &    $-$69:36:37.5    &   G0 V    &  15.7\\
23    &  5:39:30.69    &    $-$69:39:12.3    &   O8 Vz    &  14.5\\
24    &  5:39:32.37    &    $-$69:40:12.6    &   B2$-$3 V    &  16.7\\
25    &  5:39:32.66    &    $-$69:39:18.2    &   O6.5 Vz    &  14.5\\   
26    &  5:39:34.29    &    $-$69:42:50.5    &   O7 Vz    &  15.3\\   
27    &  5:39:34.90    &    $-$69:39:22.7    &   O8.5 V((f))    &  14.6\\
28    &  5:39:35.61    &    $-$69:39:12.5    &   O8$-$8.5 V    &  14.7\\
29    &  5:39:36.21    &    $-$69:37:13.7    &   O9$-$9.5 III$-$II&  15.8\\
30    &  5:39:36.51    &    $-$69:39:24.1    &   B0 V    &  15.1\\
31    &  5:39:36.86    &    $-$69:39:20.6    &   B1$-$2 V    &  16.6\\
32    &  5:39:37.17    &    $-$69:43:45.4    &   B0 Ia    &  14.4\\
33 \#    &  5:39:38.48    &    $-$69:38:23.5    &   O9.7 Iabpe  &   14.0\\
34    &  5:39:38.50    &    $-$69:36:52.3    &   G5 V    &  14.4\\
35 \#    &  5:39:38.88    &    $-$69:44:35.6    &   O8(f)p    &  14.6\\
36    &  5:39:40.04    &    $-$69:43:52.2    &   G8 V$-$III    &  13.9\\   
37 \# &  5:39:40.09    &    $-$69:46:19.6    &   O8        &  13.0\\   
38 \#    &  5:39:40.09    &    $-$69:44:33.7    &   B3 I(a)    &  12.1\\
39 \#    &  5:39:40.80    &    $-$69:38:33.7    &   O9.7 Ia+pe  &  13.0\\
40    &  5:39:41.61    &    $-$69:44:20.8    &   O6.5 V    &  14.7\\
41 \#    &  5:39:43.9    &    $-$69:38:42.9    &   O3 Vz((f*)) &  \nodata\\
42    &  5:39:44.17    &    $-$69:38:40.9    &   O6: Vz    &  \nodata\\
43    &  5:39:44.41    &    $-$69:38:40.0    &   O8$-$9 V    &  13.8\\
44    &  5:39:44.52    &    $-$69:39:12.0    &   B1$-$3 III    &  15.6\\
45    &  5:39:44.94    &    $-$69:39:00.8    &   O8$-$8.5 Vz    &  13.6\\
46    &  5:39:44.96    &    $-$69:44:21.0    &   B2$-$3 III    &  17.9\\
47    &  5:39:46.12    &    $-$69:43:57.1    &   O6 Vz((f))  &  14.3\\
48    &  5:39:46.14    &    $-$69:38:52.9    &   O4$-$6 Vz    &  13.8\\
49    &  5:39:46.80    &    $-$69:39:12.5    &   O6 V    &  14.8\\
50    &  5:39:49.67    &    $-$69:49:19.8    &   B1$-$2:    &  16.5\\
51    &  5:39:50.07    &    $-$69:39:35.9    &   O9 V    &  14.8\\
52    &  5:39:50.25    &    $-$69:43:54.0    &   B1.5$-$2 V    &  15.2\\
53    &  5:39:50.40    &    $-$69:38:18.3    &   O6: V    &  15.3\\
54 \#    &  5:39:51.13    &    $-$69:44:22.2    &   O4$-$6 Vn SB2 ? &  14.4\\
55    &  5:39:52.47    &    $-$69:40:38.9    &   B0.5$-$1 V    &  14.4\\
56    &  5:39:52.69    &    $-$69:44:05.4    &   O9.5 V$-$III&      15.8\\
57    &  5:39:52.74    &    $-$69:45:46.8    &   O8        &      14.4\\
58    &  5:39:54.30    &    $-$69:39:33.9    &   O7.5 III((f))&  14.4\\
59    &  5:39:55.00    &    $-$69:44:25.8    &   O6 III(f)    &      14.3\\
60    &  5:39:55.89    &    $-$69:39:14.9    &   B1 V    &      16.4\\
61    &  5:39:56.41    &    $-$69:38:59.6    &   O9$-$9.5 V    &      15.1\\
62    &  5:39:57.09    &    $-$69:43:53.3    &   B1$-$2 IV$-$II &   17.8\\
63    &  5:39:57.35    &    $-$69:38:59.4    &   O9 V    &      15.5\\
64    &  5:39:57.63    &    $-$69:39:14.8    &   B1 V    &      15.3\\
65    &  5:39:57.94    &    $-$69:46:03.2    &   B2 V    &      16.5\\
66    &  5:39:58.74    &    $-$69:44:04.1    &   O8.5 II((f))&      12.5\\
67    &  5:39:58.93    &    $-$69:39:25.4    &   O7 Vn    &      13.9\\
68    &  5:39:58.95    &    $-$69:43:51.8    &   O7 Vz    &      15.2\\
69    &  5:39:59.28    &    $-$69:43:57.7    &   O9.5 IV$-$III &    14.5\\
70 \#    &  5:39:59.29    &    $-$69:40:19.3    &   O9 Vn SB2 ?    &      14.7\\
71    &  5:39:59.30    &    $-$69:40:53.3    &   O9.5 V    &      15.5\\
72 \#    &  5:39:59.81    &    $-$69:36:10.6    &   O6n(f+)p    &      13.7\\
73    &  5:40:00.28    &    $-$69:40:21.7    &   O6: III:    &      13.2\\
74    &  5:40:00.50    &    $-$69:42:14.6    &   F0 Ia    &      14.3\\
75    &  5:40:00.88    &    $-$69:41:09.8    &   O9 V    &      15.1\\
76 \#    &  5:40:01.05    &    $-$69:34:52.8    &   A I+B ?    &      12.7\\
77    &  5:40:01.1    &    $-$69:46:02.5    &   O7 f/(f)    &      14.0\\
78    &  5:40:02.38    &    $-$69:40:59.1    &   B3$-$5 III    &      16.8\\   
79    &  5:40:03.06    &    $-$69:43:52.1    &   B8$-$9 V$-$III &   17.2\\
80    &  5:40:03.23    &    $-$69:40:58.7    &   B3$-$5 V$-$III &   17.1\\
81    &  5:40:04.07    &    $-$69:43:51.7    &   O7n        &      14.2\\
82 \#    &  5:40:04.62    &    $-$69:39:50.6    &   O5n(f+)p    &      12.4\\
83    &  5:40:04.68     &    $-$69:39:20.0    &   B1$-$1.5 V    &      16.9\\
84    &  5:40:04.85    &    $-$69:40:59.1    &   B1.5$-$2 II    &      14.5\\   
85    &  5:40:05.54    &    $-$69:39:18.9    &   B1.5 II    &      14.9\\
86    &  5:40:05.59    &    $-$69:46:05.4    &   B0.5$-$1 III&      15.1\\
87    &  5:40:06.83    &    $-$69:45:42.9    &   O7 V:    &      13.9\\
88    &  5:40:08.18    &    $-$69:39:17.2    &   O4 III(f)    &      13.6\\
89    &  5:40:09.07    &    $-$69:47:00.1    &   O9        &      15.9\\
90    &  5:40:08.83    &    $-$69:40:22.1    &   O6: Vzn    &      15.1\\
91    &  5:40:09.45    &    $-$69:40:18.0    &   O8$-$8.5 V    &      13.6\\   
92    &  5:40:10.56    &    $-$69:40:17.8    &   B1 V    &      14.2\\
93 \#    &  5:40:12.80    &    $-$69:34:54.6    &   BN0 Iabp    &      12.5\\
94    &  5:40:13.43    &    $-$69:34:45.2    &   O9.7 Ib    &      12.8\\
95    &  5:40:13.48    &    $-$69:43:50.3    &   K5 V    &      14.5\\
96    &  5:40:14.04    &    $-$69:38:06.7    &   O9.5 V(n)    &      14.8\\
97    &  5:40:14.26    &    $-$69:42:59.3    &   B0.2$-$0.5 V&      15.1\\
98    &  5:40:15.60    &    $-$69:36:57.5    &   O8 II((f))    &      13.9\\
99    &  5:40:17.45    &    $-$69:46:24.1    &   O6:        &      14.8\\
100    &  5:40:17.85    &    $-$69:38:02.4    &   G5 V    &      16.2\\
101    &  5:40:17.91    &    $-$69:37:06.8    &   O6 Vz((f))    &      14.2\\
102    &  5:40:18.32    &    $-$69:36:33.4    &   O8.5$-$9 III&      13.9\\
103    &  5:40:18.36    &    $-$69:37:17.3    &   B2.5 V    &      15.4\\
104    &  5:40:19.32    &    $-$69:37:07.0    &   O8.5 III-II((f))&  13.8\\
105    &  5:40:19.64    &    $-$69:40:07.5    &   O9.5 V    &      15.6\\     
106    &  5:40:20.21    &    $-$69:40:31.8    &   O8$-$8.5 Vz &      15.3\\
107 \#    &  5:40:20.58    &    $-$69:39:01.0    &   O6 Vn SB2 ?    &      14.5\\
108    &  5:40:20.90    &    $-$69:40:19.1    &   B0$-$0.2 V    &      15.3\\   
109    &  5:40:20.93    &    $-$69:37:18.6    &   B5 III$-$I    &      16.2\\
110    &  5:40:23.64    &    $-$69:37:43.4    &   B5$-$8 V$-$III &   18.1\\
111    &  5:40:24.75    &    $-$69:40:13.2    &   O6: Vz    &      12.4\\    
112 \#    &  5:40:24.78    &    $-$69:37:44.0    &   B2$-$2.5 SB2 ?&      15.3\\
113    &  5:40:25.00   &    $-$69:41:59.0    &   B0.5$-$1 V  &      15.3\\
114    &  5:40:25.20    &    $-$69:34:57.5    &   O9.5 Ib    &      13.8\\
115    &  5:40:25.98    &    $-$69:35:12.3    &   B1 Ib    &      14.0\\
116    &  5:40:26.18    &    $-$69:41:32.4    &   O6.5 Vz    &      15.0\\
117    &  5:40:28.03    &    $-$69:36:12.2    &   O6.5 Vz    &      14.3\\ 
118    &  5:40:28.19    &    $-$69:35:13.7    &   O6$-$7:    &      16.3\\
119    &  5:40:28.38    &    $-$69:43:37.6    &   O8$-$9 V    &      14.9\\
120    &  5:40:29.68    &    $-$69:45:16.7    &   B0 V      &      15.9\\
121    &  5:40:30.46    &    $-$69:37:02.3    &   B1$-$1.5 III$-$II &      15.1\\
122    &  5:40:31.32    &    $-$69:35:02.8    &   B2 V      &      15.6\\
123    &  5:40:33.42    &    $-$69:35:04.8    &   B1$-$1.5 III$-$II &      14.3\\
124    &  5:40:34.77    &    $-$69:39:40.3    &   B4 III$-$I      &      13.1\\
125    &  5:40:35.53    &    $-$69:36:10.4    &   B5 V      &      18.1\\
126    &  5:40:37.28    &    $-$69:35:23.0    &   B0.7 Ib      &      13.8\\
127    &  5:40:38.58    &    $-$69:34:25.4    &   B2 V      &      15.1\\
128    &  5:40:39.62    &    $-$69:44:32.2    &   K0 V      &      15.4\\
129    &  5:40:40.30    &    $-$69:36:42.9    &   B0$-$1 V      &      15.6\\
130    &  5:40:41.56    &    $-$69:37:33.0    &   O8 Vz      &      14.6\\
131    &  5:40:41.97    &    $-$69:45:12.4    &   F3 V      &      13.3\\
132    &  5:40:44.96    &    $-$69:38:42.3    &   F7 V      &      12.8\\
133    &  5:40:48.14    &    $-$69:43:17.6    &   O6 V      &      15.2\\
134    &  5:40:48.33    &    $-$69:39:46.8    &   B1 III      &      15.0\\
135    &  5:40:48.45    &    $-$69:49:38.6    &   B          &      14.1\\
136    &  5:40:49.51    &    $-$69:42:41.9    &   B2 II      &      14.6\\
137    &  5:40:50.26    &    $-$69:38:41.8    &   B2 V      &      14.8\\
138    &  5:40:52.82    &    $-$69:40:14.0    &   B2 V      &      17.0\\
139    &  5:40:53.57    &    $-$69:43:04.9    &   G8 V      &      14.0\\
140    &  5:40:53.69    &    $-$69:40:16.9    &   B0.2$-$0.5 V  &      15.2\\
141    &  5:40:53.72    &    $-$69:40:13.1    &   B0.2$-$0.5 V  &      15.6\\
142    &  5:40:53.72    &    $-$69:47:43.2    &   G0 V      &      14.8\\
143    &  5:40:54.86    &    $-$69:40:37.5    &   B1$-$1.5 V      &      14.6\\
144    &  5:40:55.09    &    $-$69:42:39.4    &   A0 III$-$Ib      &      14.6\\
145    &  5:40:55.77    &    $-$69:39:13.5    &   O8.5 Ib(f)      &      14.6\\
146    &  5:40:58.02    &    $-$69:48:31.2    &   B0$-$0.2 V$-$IV&      14.3\\
147    &  5:40:59.75    &    $-$69:38:40.0    &   BC1.5 Iab      &      12.2\\
148    &  5:41:01.21    &    $-$69:37:36.8    &   B1 V      &      15.1\\
149    &  5:41:04.74    &    $-$69:37:48.6    &   B1$-$1.5 V      &      14.8\\
150    &  5:41:06.48    &    $-$69:40:21.6    &   B1.5 Iab      &      13.5\\
151 \#    &  5:41:09.77    &    $-$69:39:15.8    &   O7n(f)p      &      13.2\\
152    &  5:41:11.29    &    $-$69:44:31.1    &   O9 III      &      13.9\\
153    &  5:41:15.07    &    $-$69:36:17.0    &   B2 V      &      15.9\\
154    &  5:41:16.18    &    $-$69:32:31.9    &   B7$-$8 V      &      17.6\\
155    &  5:41:16.62    &    $-$69:36:18.1    &   B1 V       &      17.1\\
156    &  5:41:17.12    &    $-$69:32:33.1    &   B0.5$-$1 V      &      14.8\\
157    &  5:41:20.02    &    $-$69:42:50.5    &   F0 V      &      12.9\\
158    &  5:41:20.04    &    $-$69:36:22.4    &   B4 III$-$I      &      11.8\\
159    &  5:41:20.24    &    $-$69:35:59.8    &   B1 V      &      14.8\\
160    &  5:41:21.40    &    $-$69:38:49.4    &   O9.5 III      &      14.5\\
161    &  5:41:22.86    &    $-$69:47:19.4    &   G0 V      &      16.3\\
162    &  5:41:26.93    &    $-$69:36:11.4    &   B1$-$2 III$-$I&      13.1\\
163    &  5:41:27.65    &    $-$69:48:03.0    &   B2.5: Ia:      &      17.2\\
164    &  5:41:27.86    &    $-$69:36:08.4    &   B2 V      &      16.0\\
165    &  5:41:28.05    &    $-$69:38:06.4    &   B1.5$-$2 III$-$II &      12.8\\
166    &  5:41:29.43    &    $-$69:46:21.8    &   O9.7 II$-$Ib  &      12.3\\
167    &  5:41:30.44    &    $-$69:36:18.7    &   B1 II      &      14.0\\
168    &  5:41:30.66    &    $-$69:40:07.4    &   O9.7 Iab      &      12.6\\
169    &  5:41:30.97    &    $-$69:39:59.7    &   B1 IV      &      16.2\\
170    &  5:41:32.68    &    $-$69:40:01.9    &   B1 III$-$II      &      13.7\\
171    &  5:41:33.22    &    $-$69:39:44.2    &   BC1 Ia      &      13.5\\
172    &  5:41:35.30    &    $-$69:40:19.9    &   B1 III$-$II      &      15.0\\
173    &  5:41:35.54    &    $-$69:35:57.7    &   B1 Ia      &      12.8\\
174    &  5:41:38.03    &    $-$69:36:16.4    &   B3$-$5 V$-$III&      18.4\\
175    &  5:41:38.85    &    $-$69:40:07.4    &   B0.5 IV      &      15.2\\
176    &  5:41:39.22    &    $-$69:49:24.2    &   B2: III$-$II: &      12.7\\
177    &  5:41:40.08    &    $-$69:39:50.8    &   B1.5 Ib      &      13.7\\
178    &  5:41:40.98    &    $-$69:39:50.8    &   O8.5$-$9 III  &      13.6\\
179 \#    &  5:41:43.71    &    $-$69:37:37.4    &   B Iae P Cyg   &      14.1\\
180    &  5:41:44.00    &    $-$69:41:24.6    &   A III        &  18.3\\
181    &  5:41:45.07    &    $-$69:40:03.5    &   F0 V     &  14.4\\
182    &  5:41:58.26    &    $-$69:39:50.1    &   A0: V$-$III     &  18.1\\
183    &  5:41:59.32    &    $-$69:43:44.1    &   A5 II$-$Ib     &  12.3\\
184    &  5:41:59.76    &    $-$69:40:04.0    &   B1$-$2 V$-$IV&  17.0\\
185    &  5:42:01.00    &    $-$69:39:48.5    &   F5$-$G0 V$-$I&  19.1\\
186    &  5:42:04:36    &    $-$69:39:48.8    &   G8 V$-$III     &  17.4\\
187    &  5:42:05.65    &    $-$69:41:23.0    &   B2$-$3 III     &  17.2\\
188    &  5:42:05.95    &    $-$69:40:04.1    &   B1$-$2 V     &  17.6\\
189    &  5:42:10.13    &    $-$69:38:45.8    &   B3$-$5 V     &  17.3\\
\enddata

\tablecomments{Column 1 lists our internal nomenclature for the observed
objects, objects are listed by R.A. Stars marked with ``\#'' have
individual comments in Section 3.2. Celestial coordinates (J2000), obtained through 2MASS astrometry, are shown in Columns 2 and 3. Spectral classification from our data is given in Column 4.
Column 5 shows, as a reference, $V$ magnitudes from the \cite{2004AJ....128.1606Z} photometric survey for all the objects, except for stars 1, 41, and 42 for which the catalog correlation was uncertain.}
\label{tab:ourspec}

\end{deluxetable}

\begin{deluxetable}{llll}
\tablecaption{Stars with spectral classification found in the
  literature.}
\tablehead{
\colhead{ID} & \colhead{Spectral Type\ (this work)} & \colhead{Sanduleak \ (ID, Spectral Type)} & \colhead{Spectral Type\ (Other)}
}
\startdata
7    & A0 Ib         & Sk $-$69 242, A2 I  & A7  (HD)          \\
     &               &                     & A2 Ia0 (A$-$P)    \\
     &               &                     & A0 Ia (S)         \\
35   &   O8(f)p      &   \nodata           & LMC~X-1 (HC)      \\
     &               &                     & O7$-$9~III (MP-B) \\
37   &   O8          &    \nodata          & O3-O6 V (C)       \\ 
38   & B3 I(a)       & Sk $-$69 254, OB    & B3~I(Nstr)+neb (F)\\
     &               &                     & B5+neb (R$-$P)    \\
66   & O8.5 II((f))  & Sk $-$69 257, OB    & B0.5 (R$-$P)      \\ 
     &               &                     & O9 II (W)         \\
73   &   O6: III:    &   \nodata           & B1-2 (BI-263)     \\
82   &   O5n(f+)p    &   \nodata           & OB0 (BI-265)      \\
     &               &                     & O6 V (M)           \\
147  & BC1.5 Iab     & Sk $-$69 268, OB    & B2.5 (P)          \\
150  & B1.5 Iab        & Sk $-$69 269, OB: & B1 III: (M)       \\
158  & B4 III$-$I    & Sk $-$69 271, OB    & B2 (P)            \\
     &               &                     & HXMB (L)          \\ 
163  & B2.5: Ia:     & Sk $-$69 274, OB:   & B2 Ia (A$-$P$-$R) \\
     &               &                     & Con (HD)          \\
     &               &                     & B2.5 Ia (F)       \\
173  & B1 Ia         & Sk $-$69 275, OB    & B1.5 (P)          \\
176  & B2: III$$-$$II: & Sk $-$69 277, OB  & B1.5 (P)          \\
183  & A5 II$$-$$Ib &   \nodata           & A5 (HD)           \\
     &               &                     & A5 I (P)          \\ 
\enddata

\tablecomments{Column 1 identifies stars according to Table~\ref{tab:ourspec}. Column 2 repeats our spectral classification already given in Table~\ref{tab:ourspec}, Column 3 contains spectral types from \cite{1970CoTol..89.....S}, and Column 4 lists classification from other sources which are individually labeled.}
\tablerefs{(HD) \citealt{1918AnHar..92....1C}; (BI) \citealt{1975A&AS...21..109B}; (R) \citealt{1960MNRAS.121..337F}; (C) \citealt{1991ApJ...373..100C}; (A) \citealt{1972A&AS....6..249A}; (P) \citealt{1978A&AS...31..243R}; (F) \citealt{1991PASP..103.1123F}; (W) \citealt{1977ApJ...215...53W}; (L) \citealt{2000A&AS..147...25L}; (MP) \citealt{1984IAUS..108..317P}; (B) \citealt{1985A&A...146..242B}; (HC) \citealt{1983ApJ...275L..43H}; (M) \citealt{1995ApJ...438..188M}; (S) \citealt{1976A&AS...24...35S}.}
\label{tab:speclit}

\end{deluxetable}


\begin{figure}
\includegraphics[width=.9\textwidth]{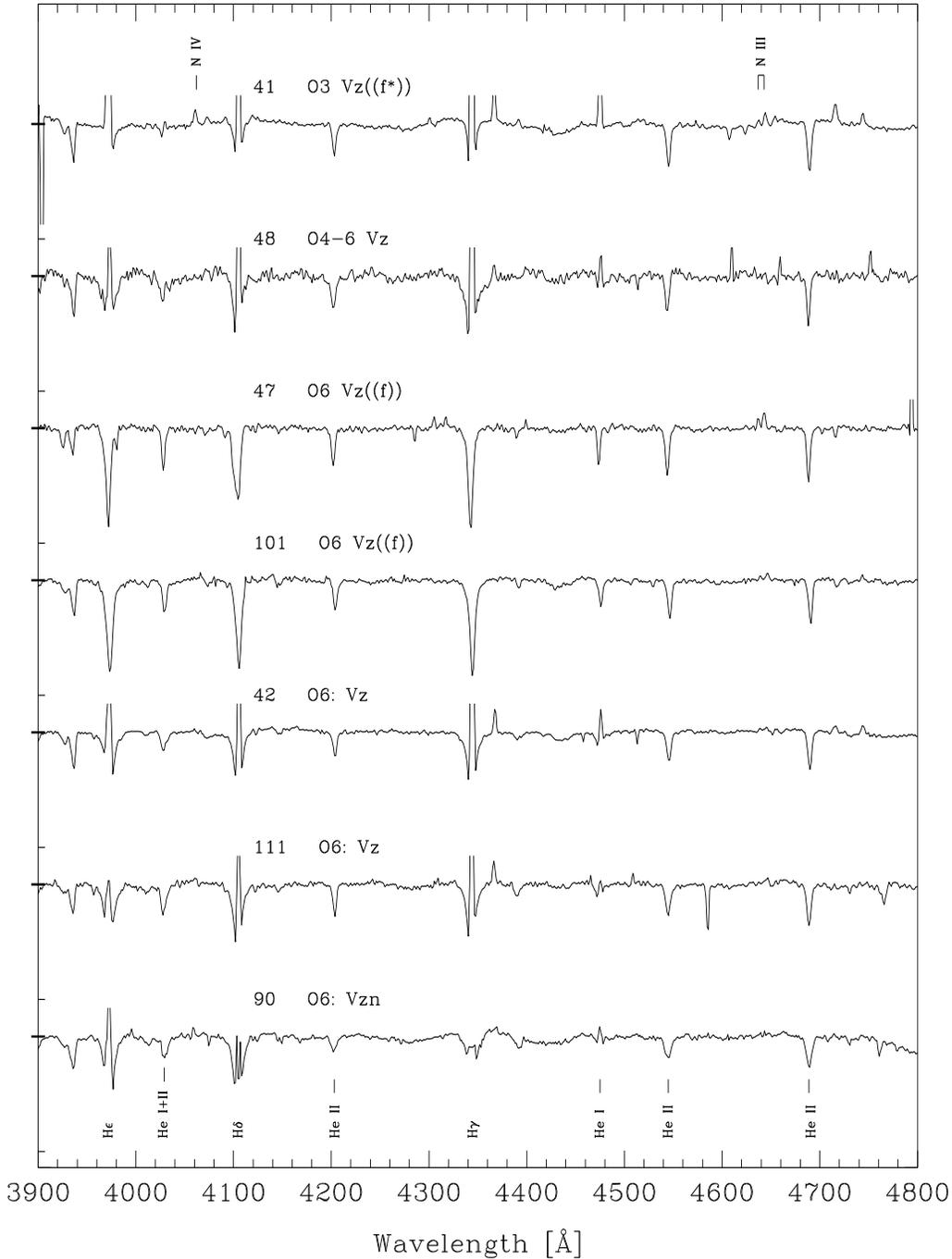}
\caption{\footnotesize  Normalized spectrograms of stars classified as early O-type
dwarfs. Wavelength in \AA\ is given on the x-axis, and on the y-axis thick long ticks mark the continuum flux while thin shorter ticks show the 0.8 continuum flux unit level. The spectral absorption features identified as reference are, from left to right by ion, He {\sc i+ii} $\lambda$4026; \heii$\lambda\lambda$4200,4541,4686; and \hei$\lambda$4471 absorption. Emission lines \niv $\lambda$4058 and \niii$\lambda\lambda$4634-40-42 are also identified above star 41.
Three Balmer series lines are also identified: H$\epsilon$ $\lambda$3970, H$\delta$ $\lambda$4101, and H$\gamma$ $\lambda$4340.}
\label{fig:01v21}
\end{figure}

\newpage

\begin{figure}
\includegraphics[width=.9\textwidth]{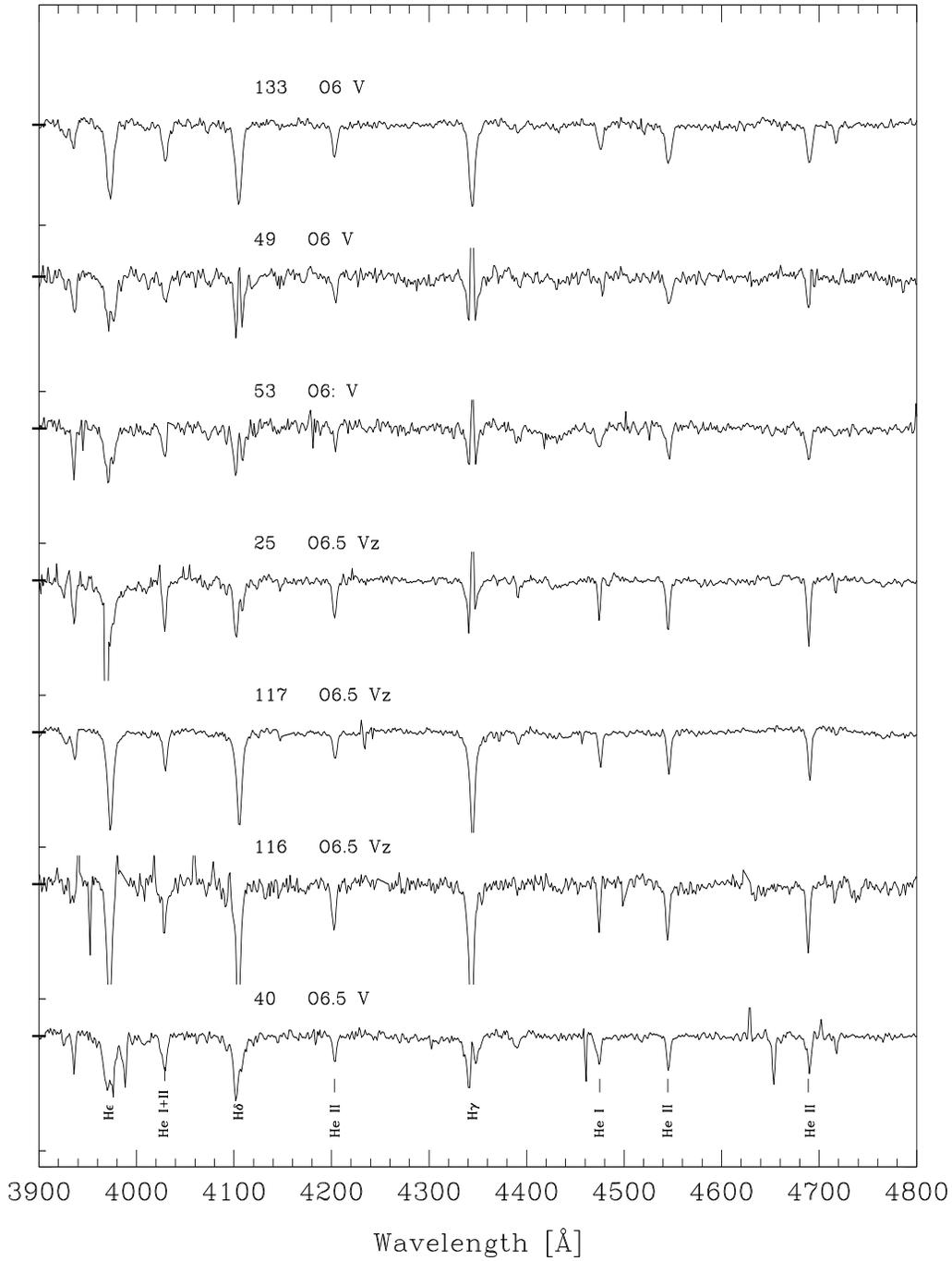}
\caption{\footnotesize Same as Figure \ref{fig:01v21} for the mid O-type dwarf stars.}
\label{fig:02v21}
\end{figure}

\newpage

\begin{figure}
\includegraphics[width=.9\textwidth]{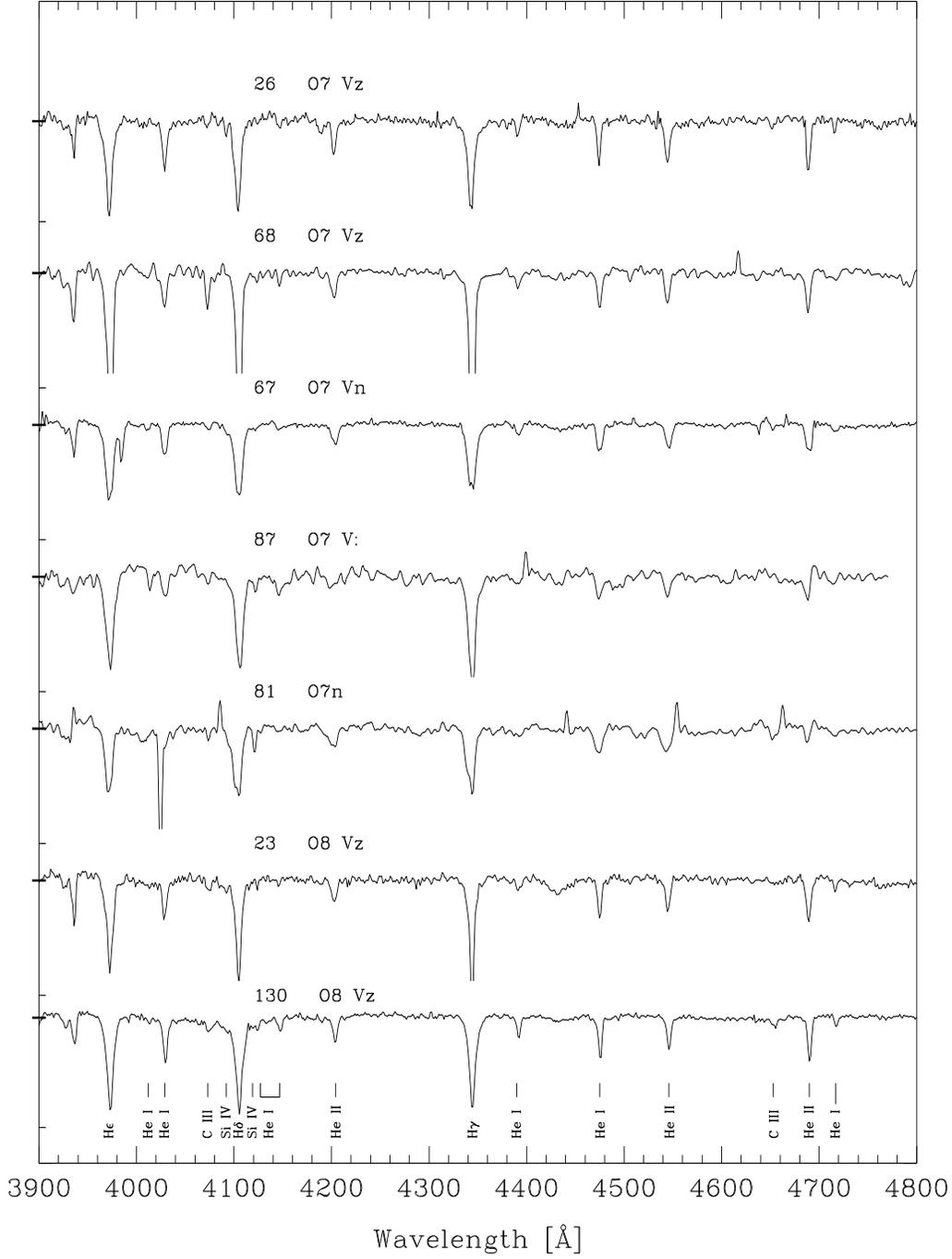}
\caption{\footnotesize Same as Figure \ref{fig:01v21} for the late O-type dwarf stars. Features identified as reference are, from left to right by ion, \hei $\lambda\lambda$4009,4026,4121,4144,4387,4471,4713; \ciii $\lambda\lambda$4070,4650 blends; \siiv $\lambda\lambda$4089,4116; and \heii $\lambda\lambda$4200,4541,4686. Three Balmer series lines are also identified: H$\epsilon$ $\lambda$3970, H$\delta$ $\lambda$4101, and H$\gamma$ $\lambda$4340.}
\label{fig:03v21}
\end{figure}

\newpage

\begin{figure}
\includegraphics[width=.9\textwidth]{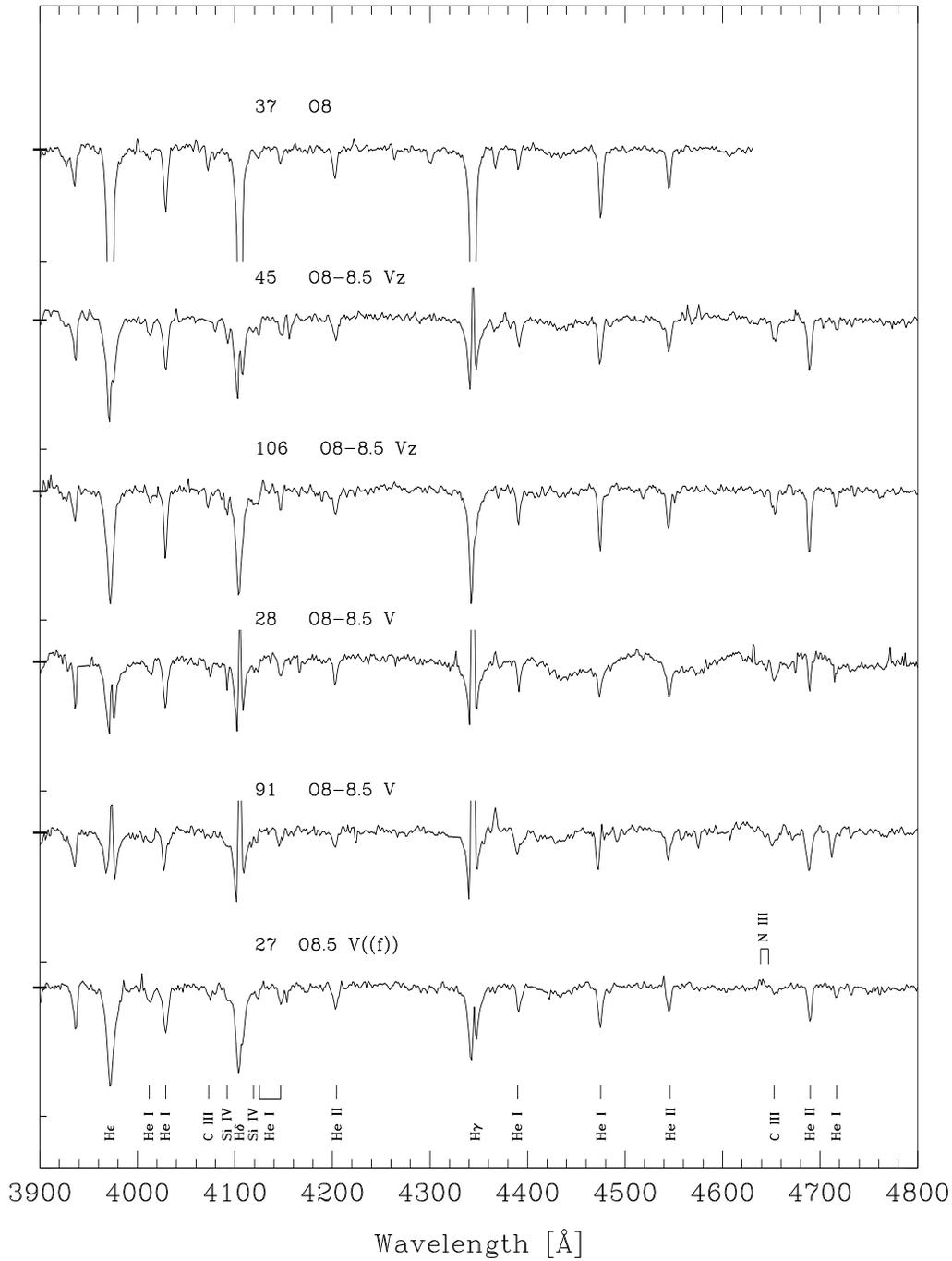}
\caption{\footnotesize Same as Figure \ref{fig:03v21} for the late O-type dwarf stars. In star 27 \niii $\lambda\lambda$4634-40-42 are also identified.}
\label{fig:04v21}
\end{figure}

\newpage

\begin{figure}
\includegraphics[width=.9\textwidth]{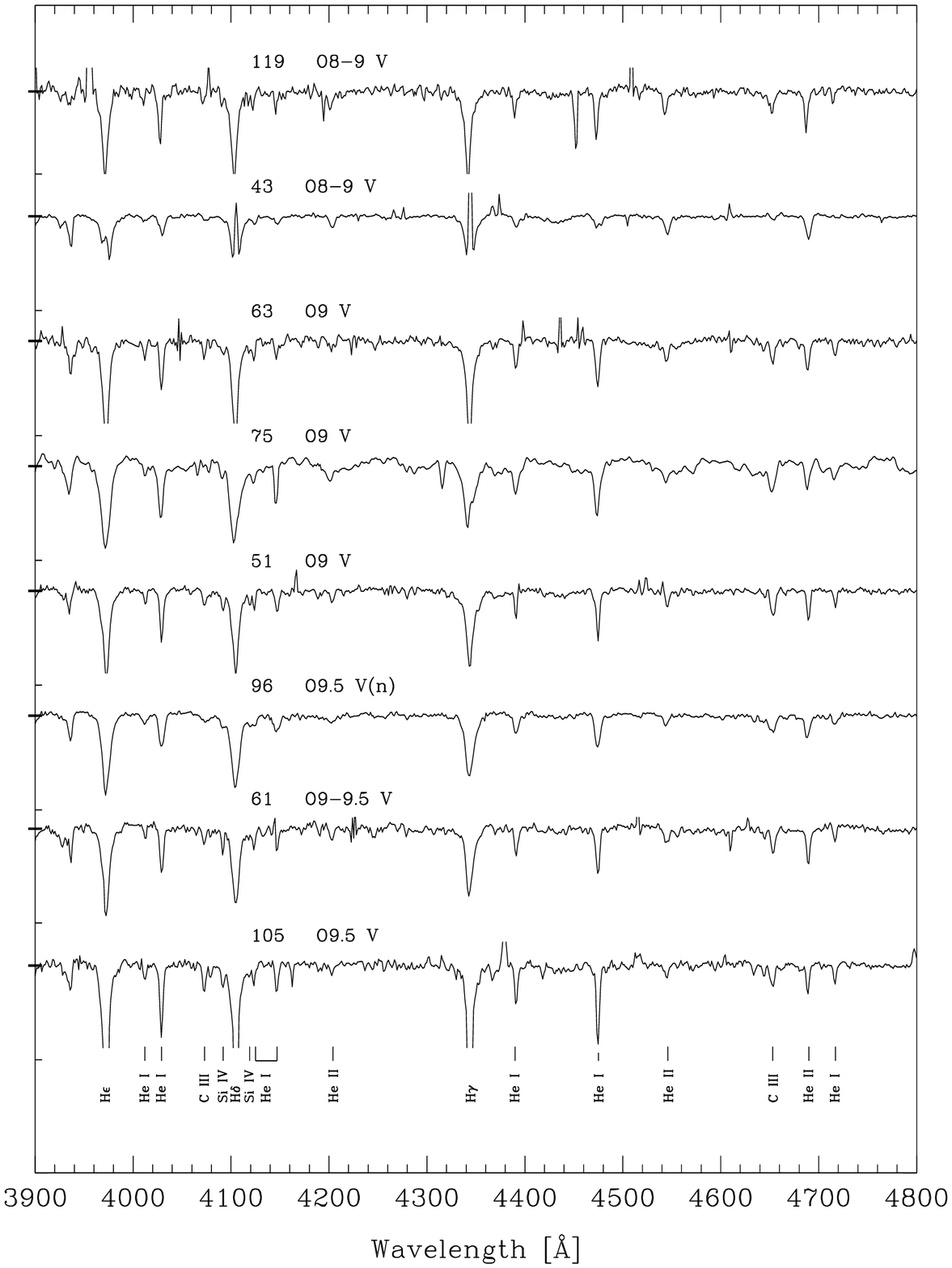}
\caption{\footnotesize Same as Figure \ref{fig:03v21}.}
\label{fig:05v21}
\end{figure}

\newpage

\begin{figure}
\includegraphics[width=.9\textwidth]{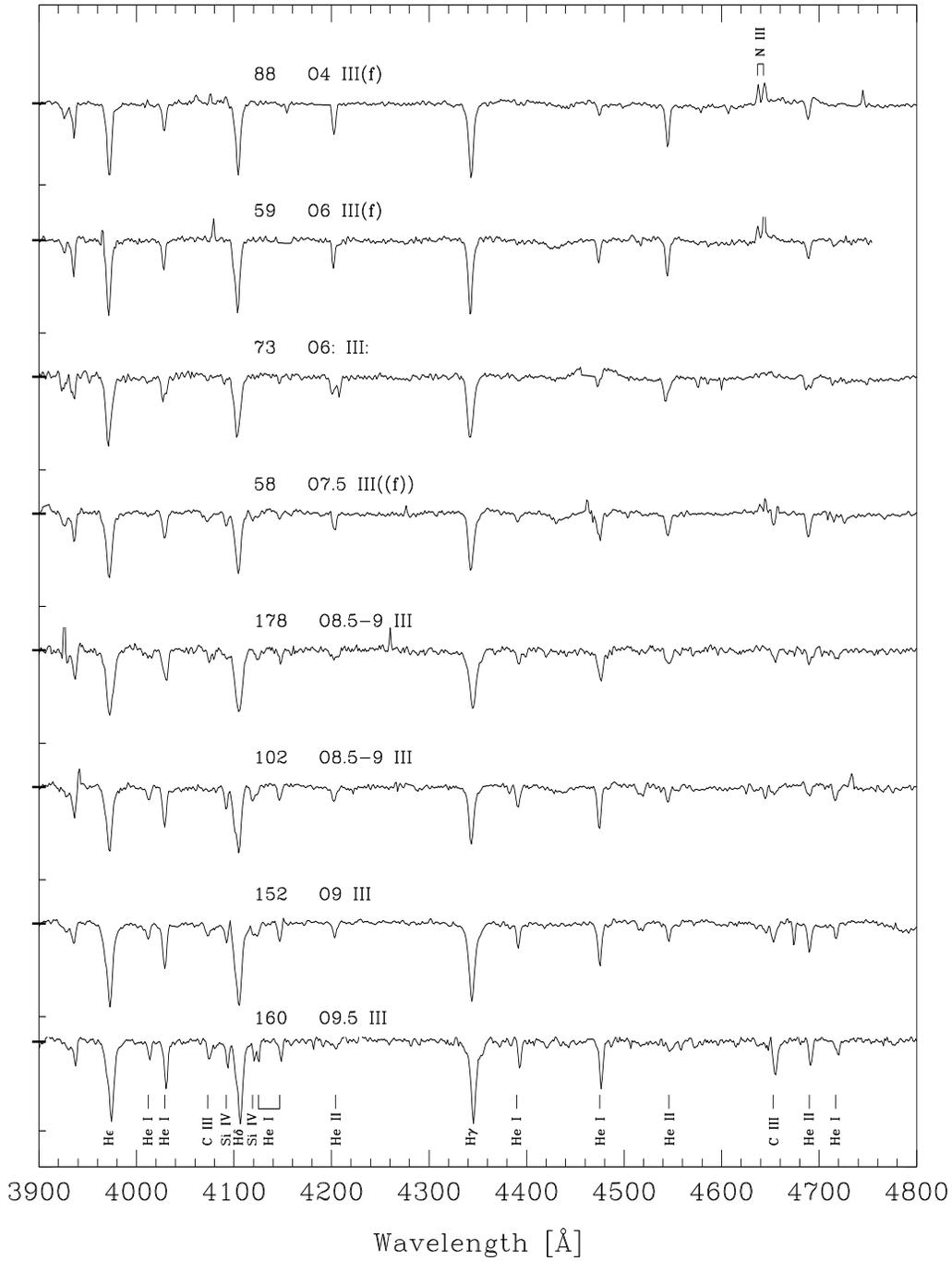}
\caption{\footnotesize O4-O9.5 giants. The spectral features identified are the same as in  Figure \ref{fig:03v21}.}
\label{fig:06v21}
\end{figure}

\newpage

\begin{figure}
\includegraphics[width=.9\textwidth]{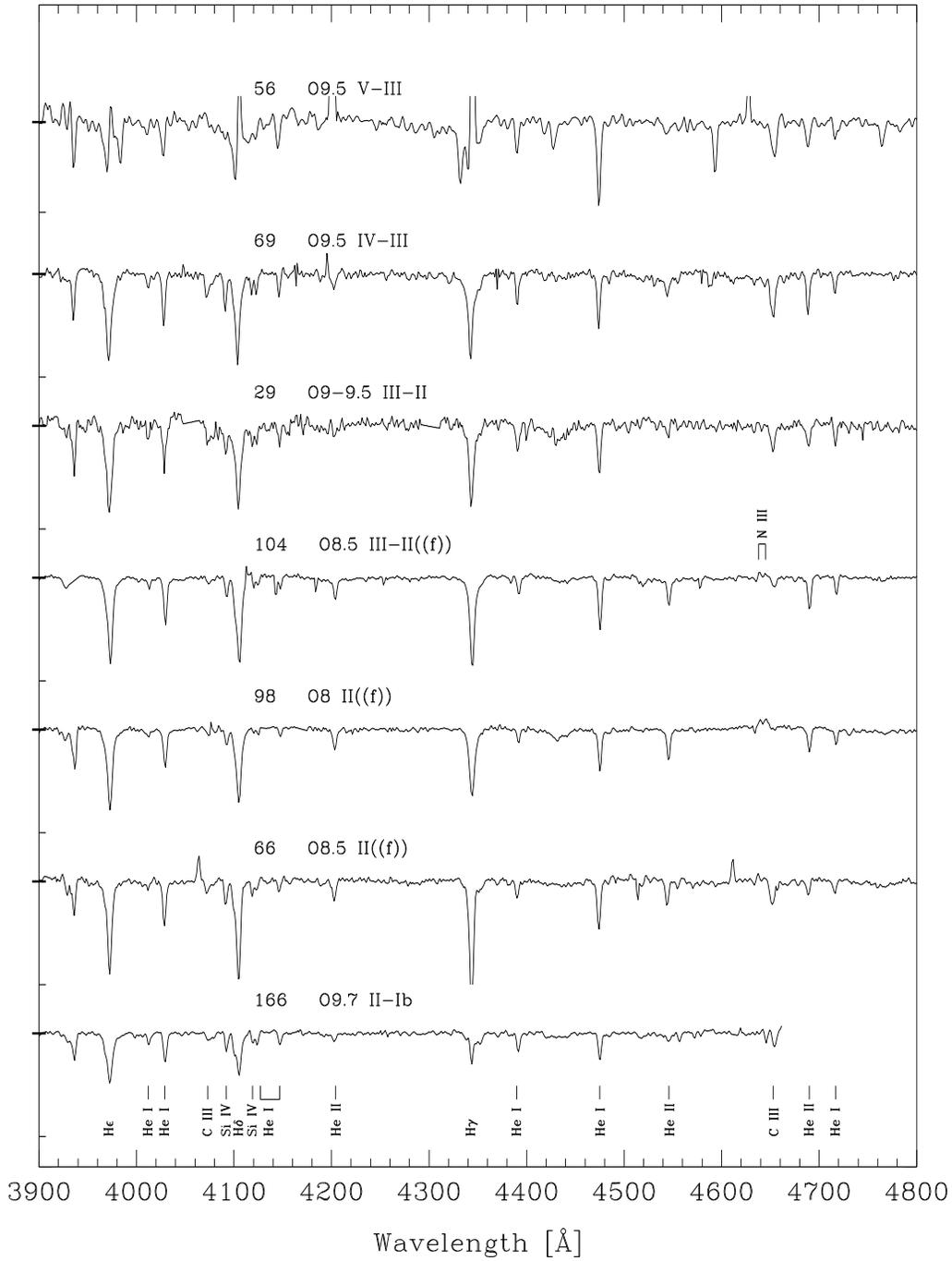}
\caption{\footnotesize Late O at luminosity classes V to Ib. The spectral features identified are the same as in  Figure \ref{fig:03v21}.}
\label{fig:07v21}
\end{figure}

\clearpage

\begin{figure}
\includegraphics[width=.9\textwidth]{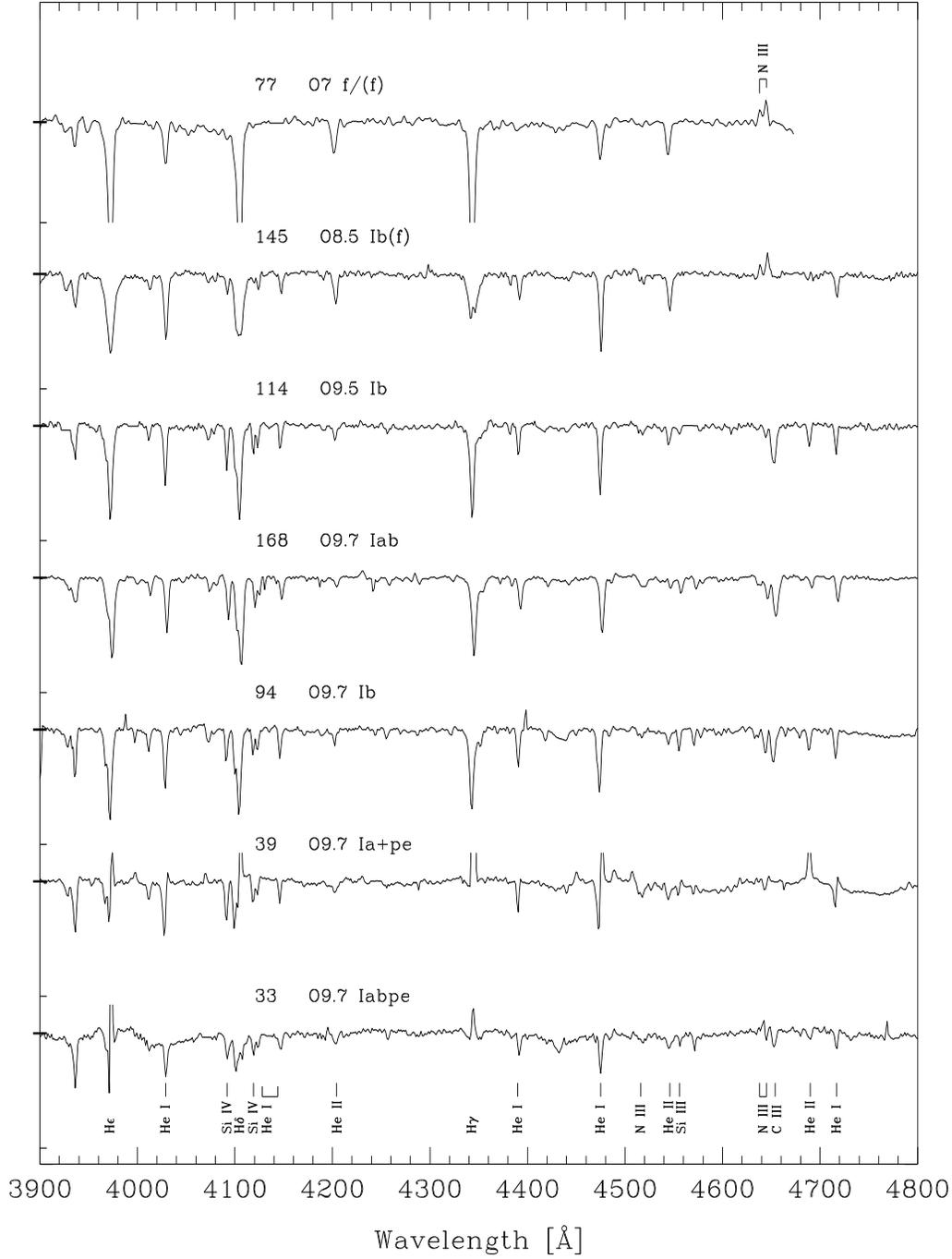}
\caption{\footnotesize Late O supergiants. The spectral features identified are, from left
  to right by ion, \hei $\lambda\lambda$4026,4121,4144,4387,4471,4713; \siiv $\lambda\lambda$4089,4116; \heii $\lambda\lambda$4200,4541,4686; \niii $\lambda\lambda$4511-15,4634-40-42; \siiii $\lambda$4552; and \ciii $\lambda$4650 blend. Three Balmer series lines are also identified: H$\epsilon$ $\lambda$3970, H$\delta$ $\lambda$4101, and H$\gamma$ $\lambda$4340.}
\label{fig:08v21}
\end{figure}

\newpage

\begin{figure}
\includegraphics[width=.9\textwidth]{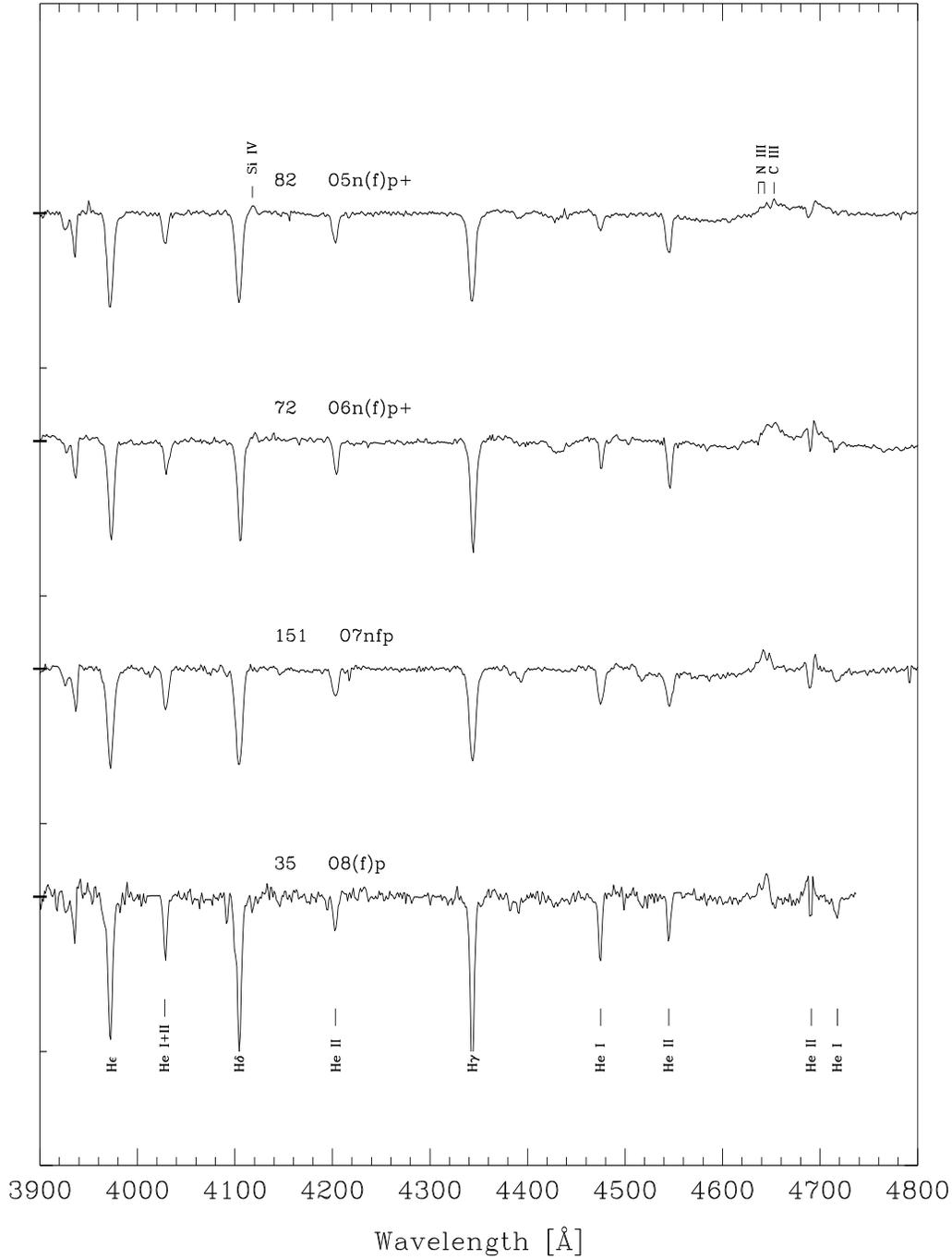}
\caption{\footnotesize O stars belonging to Onfp category. The spectral absorption features identified as reference are, from left to right by ion, He\,{\sc i}+{\sc ii} $\lambda$4026; \heii $\lambda\lambda$4200,4541,4686; \hei $\lambda\lambda$4471,4713. Emission lines identified are: \siiv $\lambda$4116; \niii $\lambda\lambda$4634-40-42; and \ciii $\lambda$4650 (see Appendix). Three Balmer series lines are also identified: H$\epsilon$ $\lambda$3970, H$\delta$ $\lambda$4101, and H$\gamma$ $\lambda$4340.}
\label{fig:09v21}
\end{figure}

\newpage

\begin{figure}
\includegraphics[width=.9\textwidth]{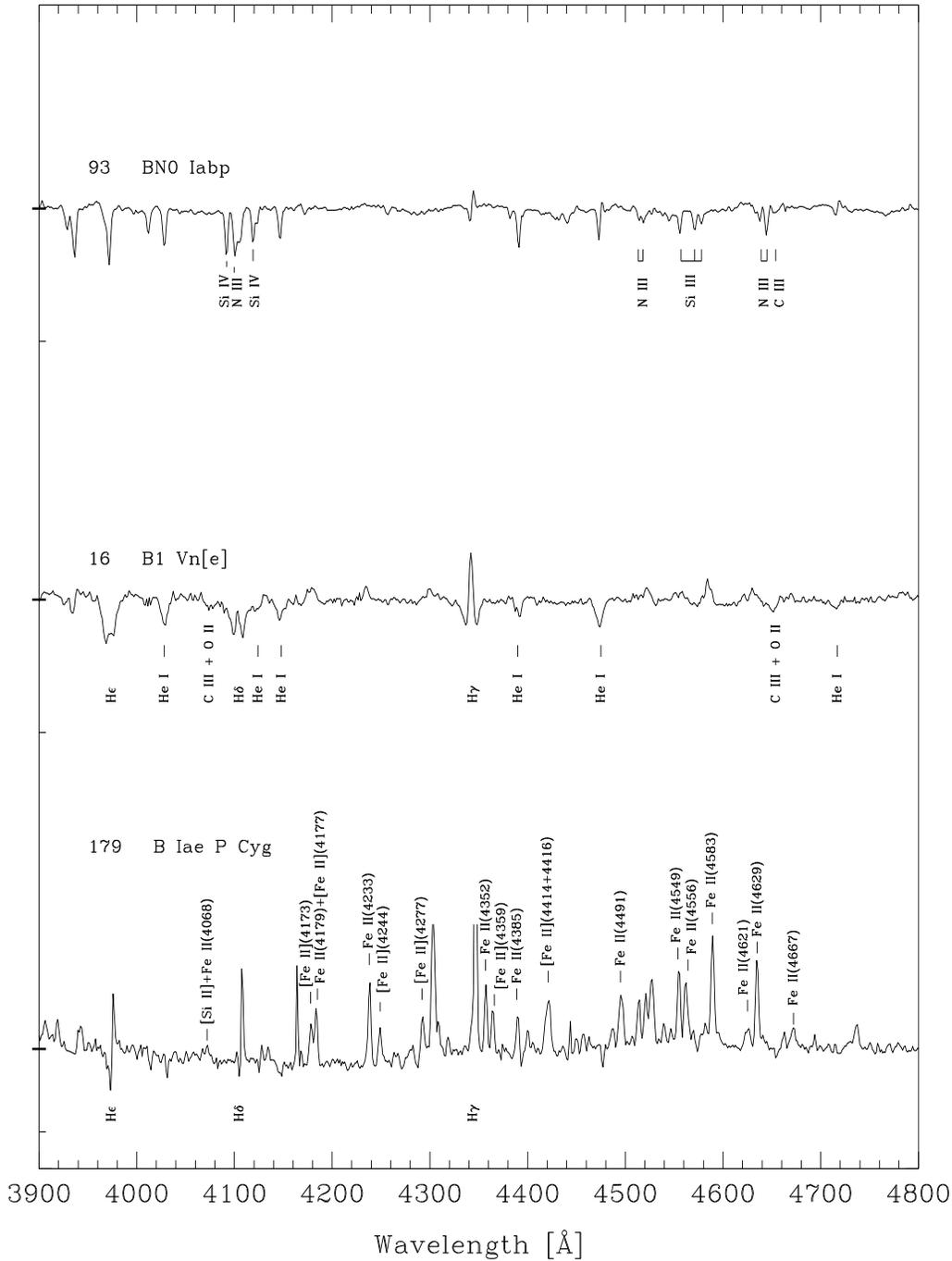}
\caption{\footnotesize Spectra of three uncommon B-type stars (See Section 3.2). The spectral features identified in star 93 are \siiv$\lambda\lambda$4089,4116; \niii$\lambda\lambda$4097,4511-15,4634-40-42 blend; \siiii$\lambda\lambda$4552-68-75; and \ciii$\lambda$4650 blend. In star 16 these are \hei$\lambda\lambda$4026,4121,4144,4387,4471,4713; C\,{\sc iii}\,+\,\oii$\lambda\lambda$4070,4650 blends; the three Balmer series lines: H$\epsilon$ $\lambda$3970, H$\delta$ $\lambda$4101, and H$\gamma$ $\lambda$4340. In star 179 the lines are identified by their respective ion and wavelengths in \AA.}
\label{fig:10v21}
\end{figure}

\newpage

\begin{figure}
\includegraphics[width=.9\textwidth]{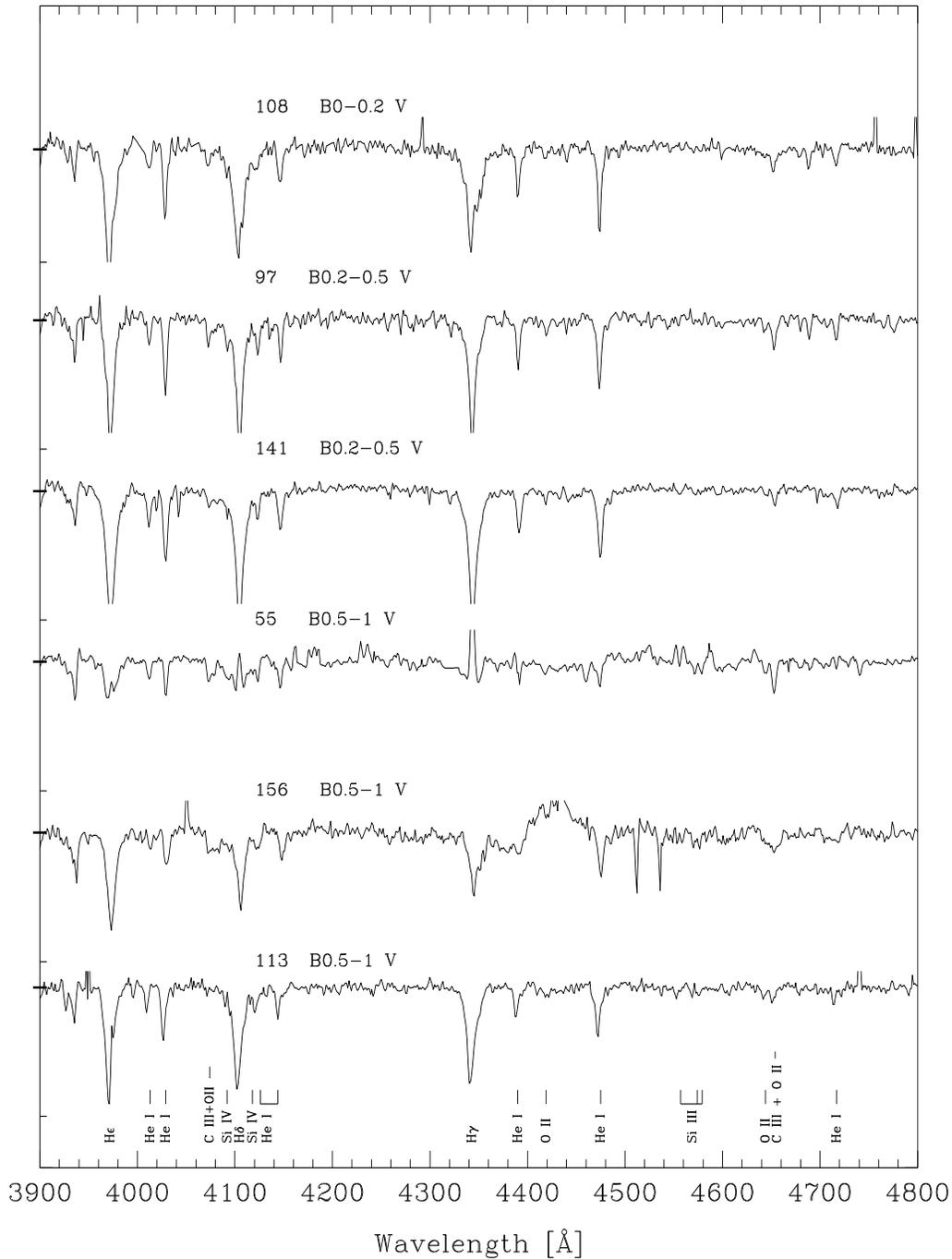}
\caption{\footnotesize Normalized spectra of stars classified as early B-type dwarf. Wavelength in \AA\ is given on the x-axis, and on the y-axis thick long ticks mark the continuum flux while thin shorter ticks show the 0.8 continuum flux unit level. The spectral absorption features identified as reference are, from left to right by ion, \hei $\lambda\lambda$4009,4026,4121,4144,4387,4471,4713; C\,{\sc iii}\,+\,\oii$\lambda\lambda$4070,4650; \siiv $\lambda\lambda$4089,4116; \oii $\lambda\lambda$4415-17,4640; and \siiii $\lambda\lambda$4552-68-75. Three Balmer series lines are also identified: H$\epsilon$ $\lambda$3970, H$\delta$ $\lambda$4101, and H$\gamma$ $\lambda$4340.}
\label{fig:11v21}
\end{figure}

\newpage

\begin{figure}
\includegraphics[width=.9\textwidth]{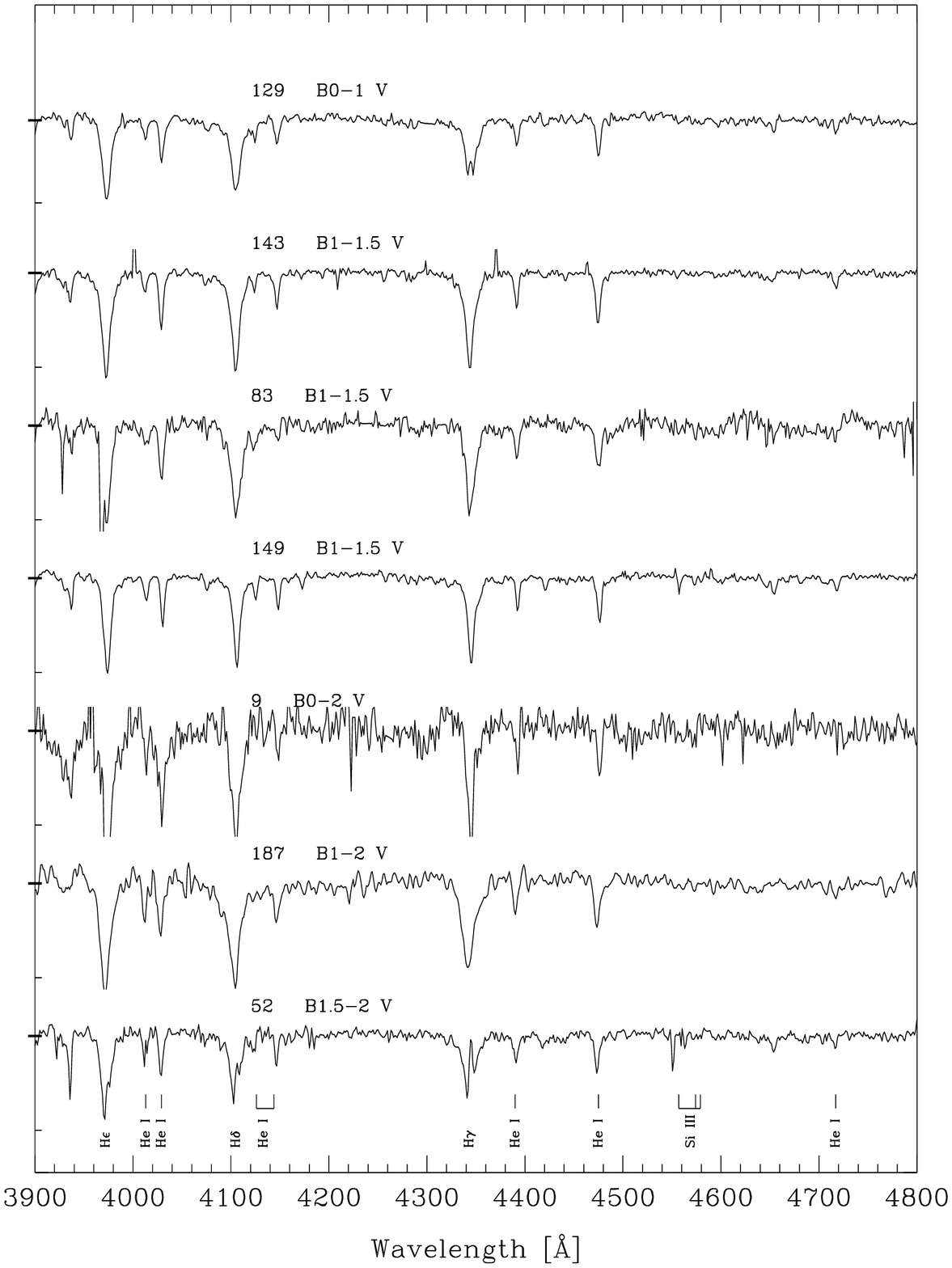}
\caption{\footnotesize Same as Figure \ref{fig:11v21}. The spectral features identified are \hei $\lambda\lambda$4009,4026,4121,4144,4387,4471,4713; \siiii $\lambda\lambda$4552-68-75; and  the Balmer series lines H$\epsilon$ $\lambda$3970, H$\delta$ $\lambda$4101, and H$\gamma$ $\lambda$4340.}
\label{fig:12v21}
\end{figure}

\newpage

\begin{figure}
\includegraphics[width=.9\textwidth]{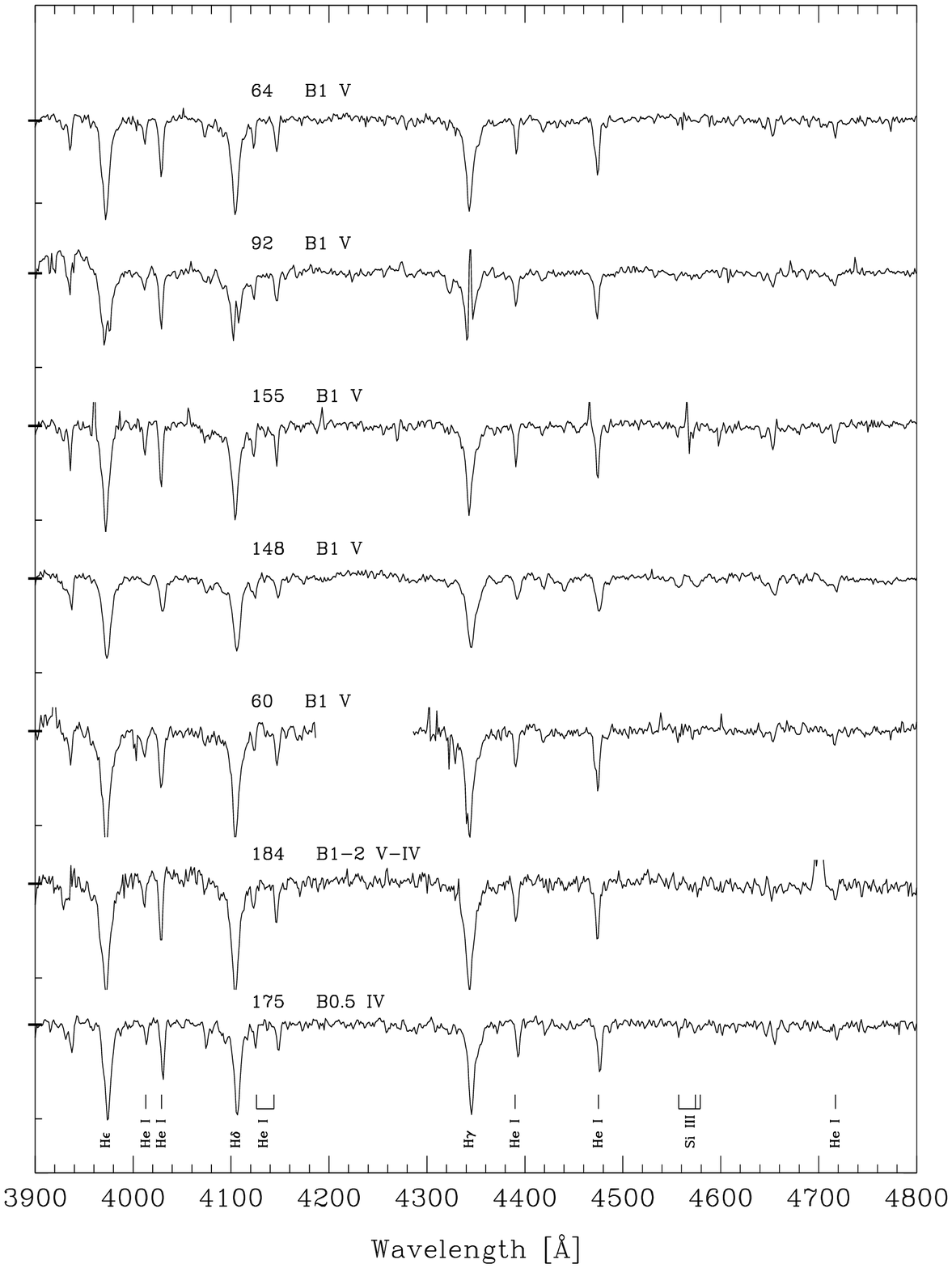}
\caption{\footnotesize Same as Figure \ref{fig:12v21}.}
\label{fig:13v21}
\end{figure}

\newpage

\begin{figure}
\includegraphics[width=.9\textwidth]{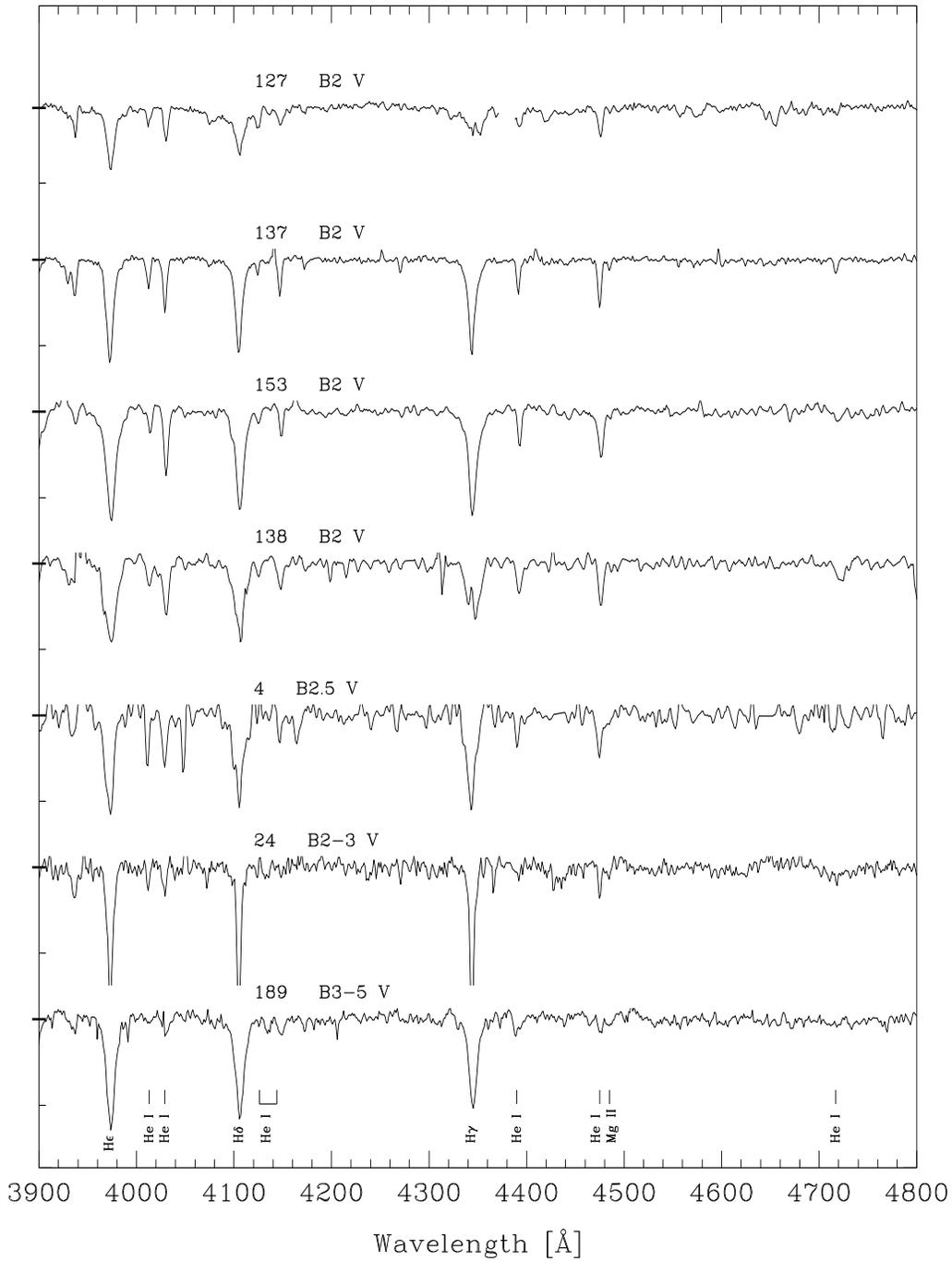}
\caption{\footnotesize Same as Figure \ref{fig:11v21}. The spectral features identified are \hei $\lambda\lambda$4009,4026,4121,4144,4387,4471,4713; \mgii $\lambda$4481; and three Balmer series lines H$\epsilon$ $\lambda$3970, H$\delta$ $\lambda$4101, and H$\gamma$ $\lambda$4340.}
\label{fig:14v21}
\end{figure}

\newpage

\begin{figure}
\includegraphics[width=.9\textwidth]{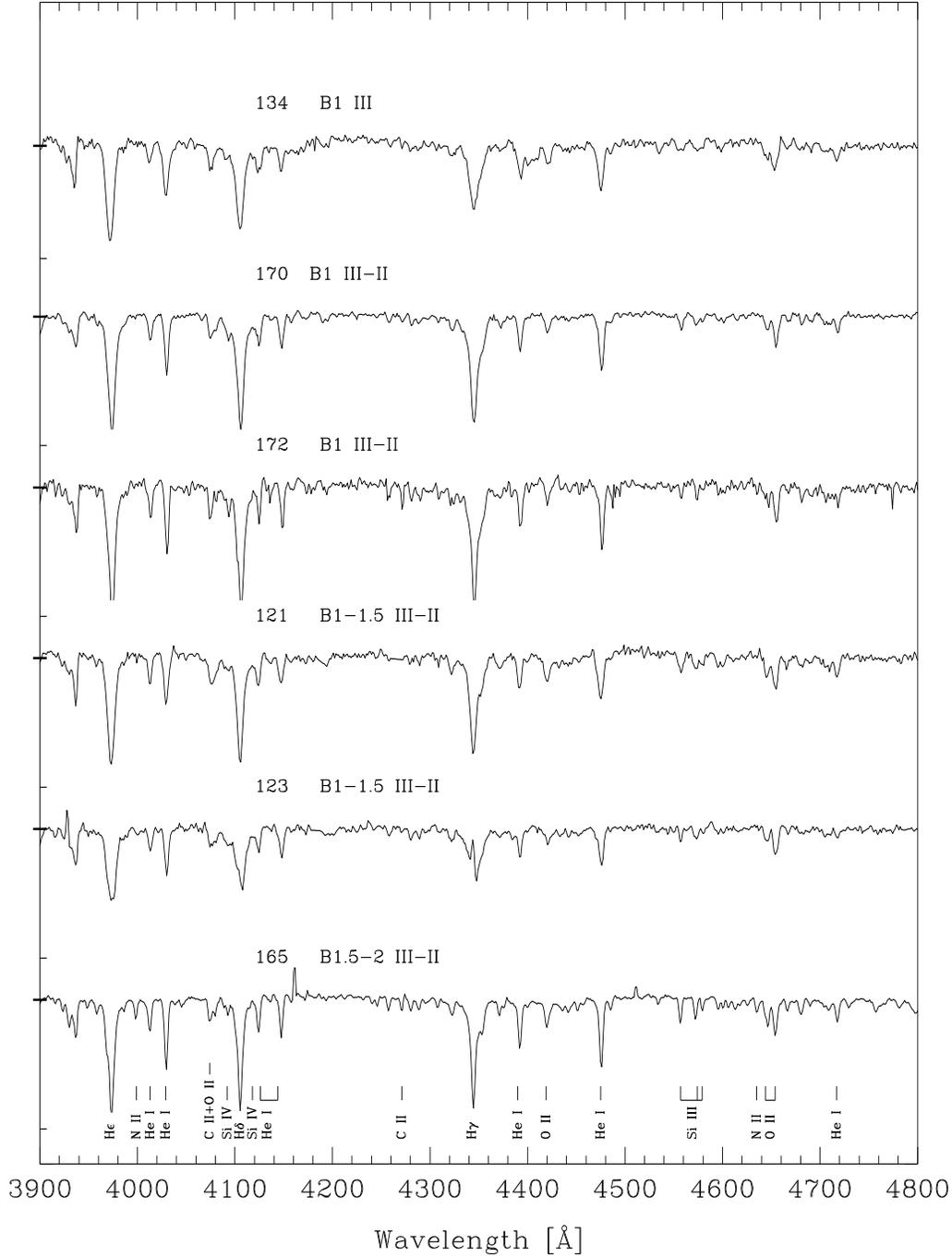}
\caption{\footnotesize Giant B stars. The spectral features identified are: \niii$\lambda\lambda$3995,4631,4643(blend+O{\sc ii}); \hei$\lambda\lambda$4009,4026,4121,4144,4387,4471,4713; C\,{\sc ii}\,+\,\oii$\lambda$4070; \siiv$\lambda\lambda$4089,4116; \cii$\lambda$4267; \oii$\lambda\lambda$4415-17,4650; and \siiii$\lambda\lambda$4552-68-75. Three Balmer series lines are also identified: H$\epsilon$ $\lambda$3970, H$\delta$ $\lambda$4101, and H$\gamma$ $\lambda$4340.} 
\label{fig:15v21}
\end{figure}

\newpage

\begin{figure}
\includegraphics[width=.9\textwidth]{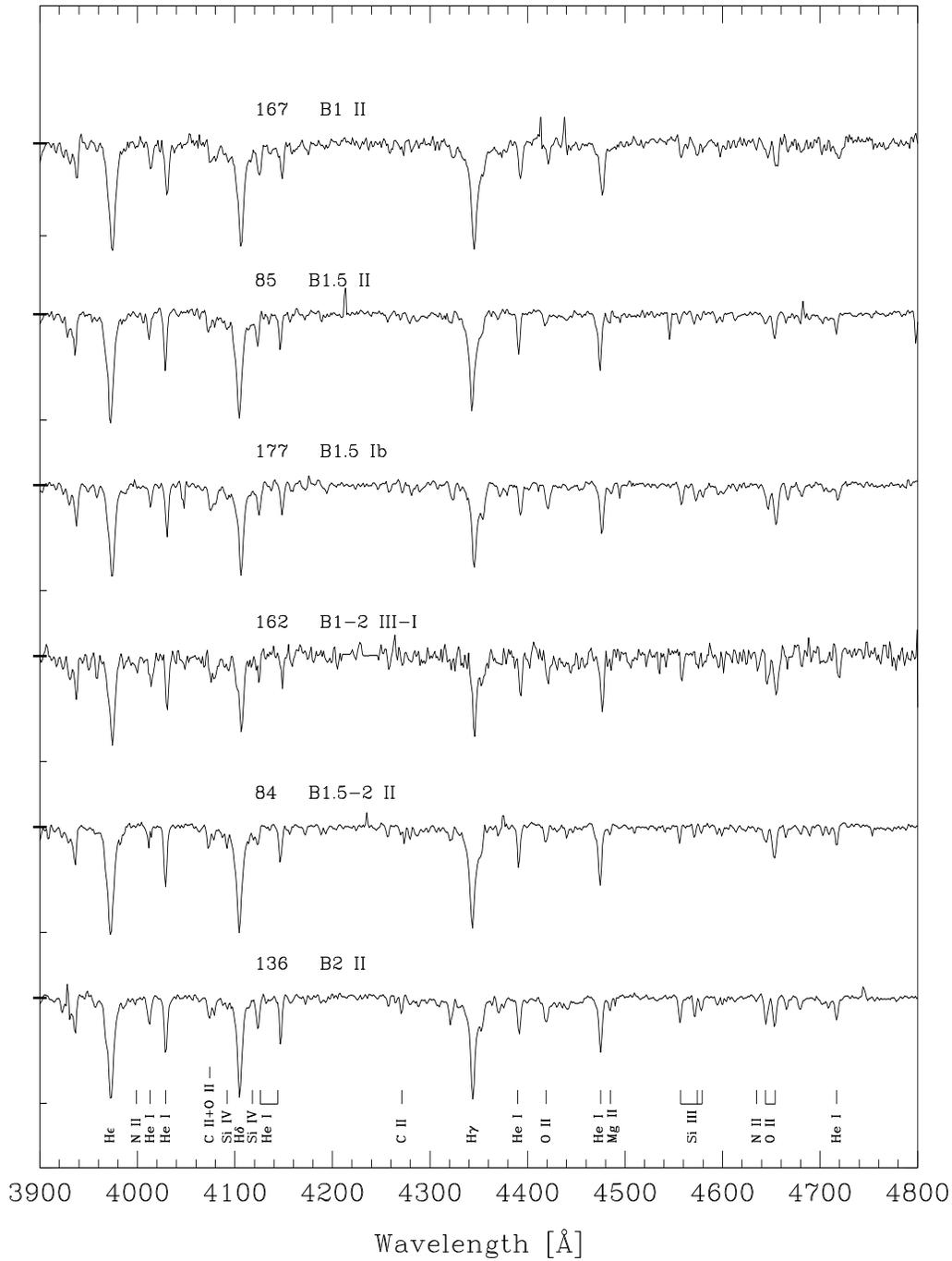}
\caption{\footnotesize Same as Figure \ref{fig:15v21} for B stars with luminosity classes ranging from giant to supergiant, \mgii $\lambda$4481 is also identified.}
\label{fig:16v21}
\end{figure}

\newpage

\begin{figure}
\includegraphics[width=.9\textwidth]{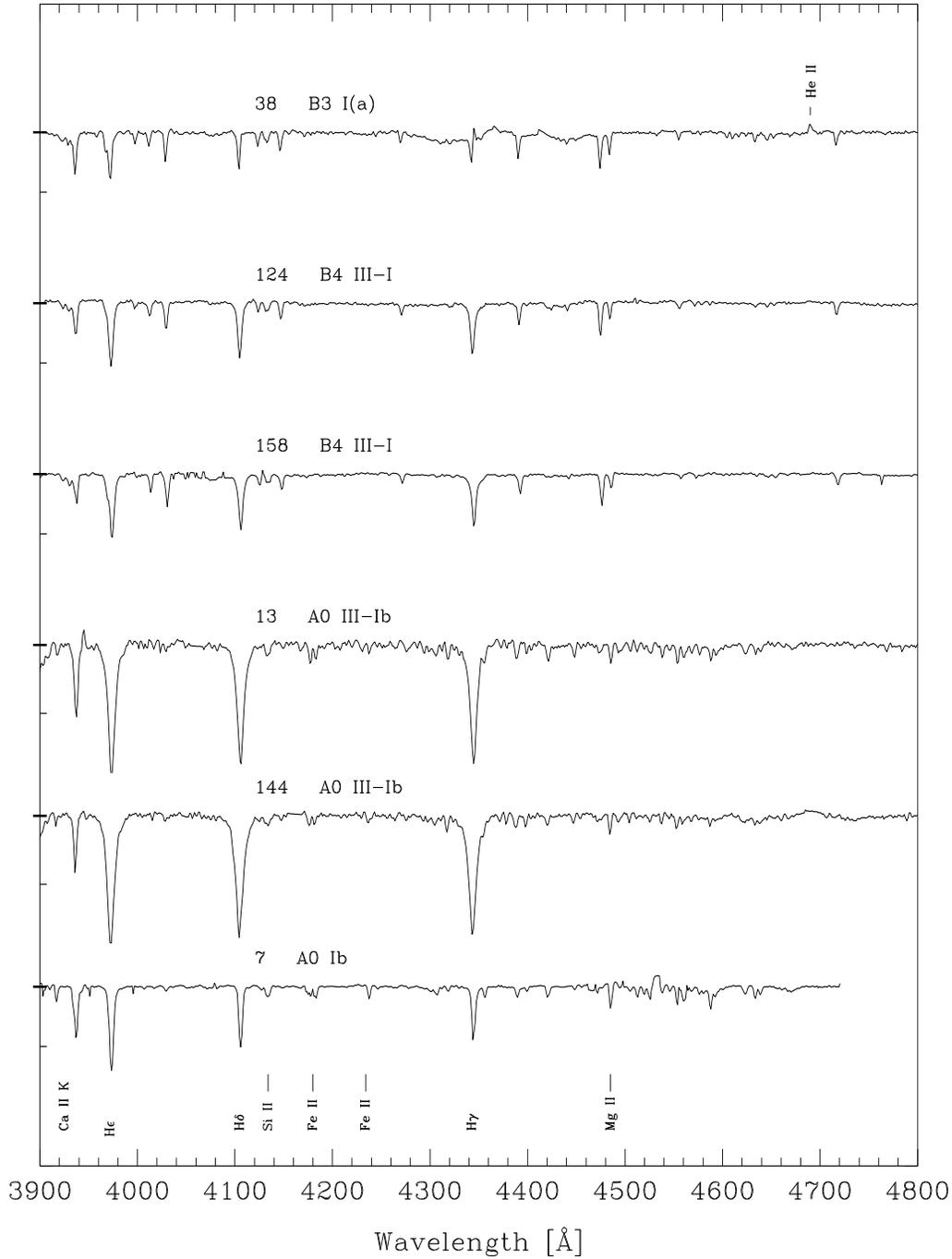}
\caption{\footnotesize Giant stars between mid B-type and mid A-type. The spectral features identified are Ca\,{\sc ii} K $\lambda$3933; \siii $\lambda\lambda$4128-30; Fe\,{\sc II} $\lambda\lambda$4172-8,4233; and \mgii $\lambda$4481.  \heii $\lambda$4686 in emission was identified in star 38 (see Section 3.2). Three Balmer series lines are also identified: H$\epsilon$ $\lambda$3970, H$\delta$ $\lambda$4101, and H$\gamma$ $\lambda$4340.}
\label{fig:17v21}
\end{figure}

\newpage

\begin{figure}
\includegraphics[width=.9\textwidth]{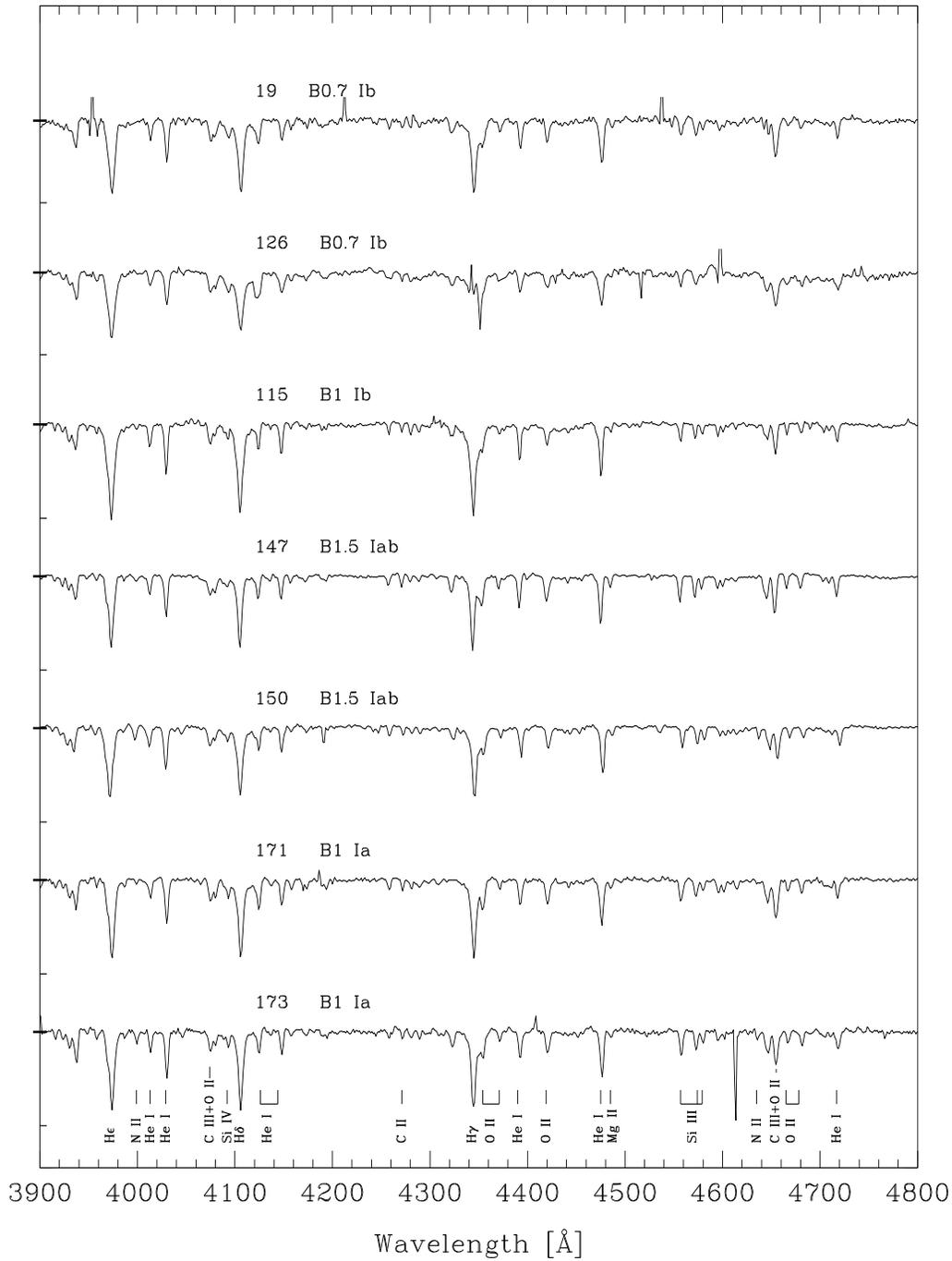}
\caption{\footnotesize B supergiant stars. Same lines as in Figure \ref{fig:15v21} are identified. \oii $\lambda\lambda$4350,4367 are also identified in star 173.}
\label{fig:18v21}
\end{figure}

\newpage

\begin{figure}
\includegraphics[width=.9\textwidth]{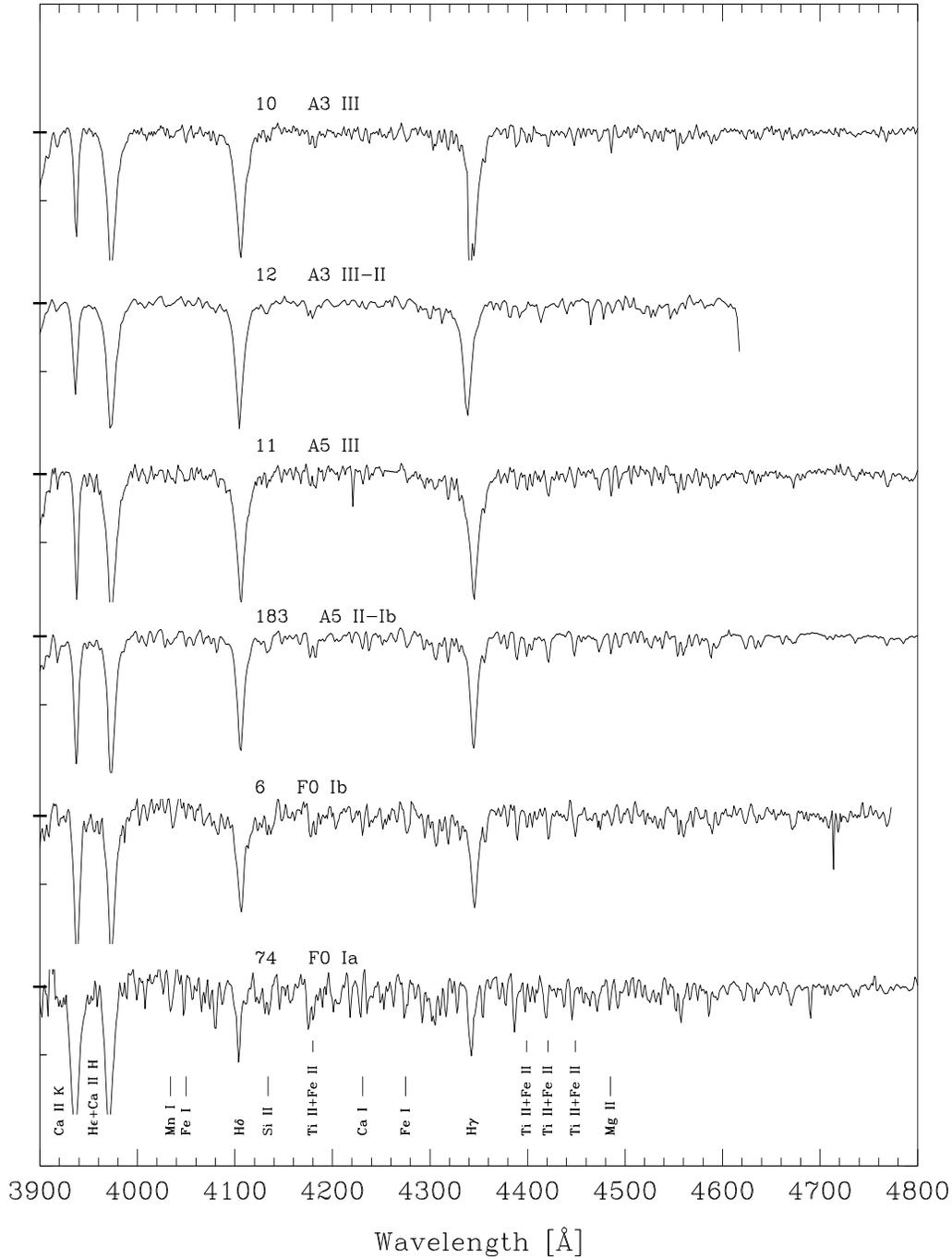}
\caption{\footnotesize From giant A3 to F0 supergiant stars. Lines identified in star 74 are Ca\,{\sc
    ii} K $\lambda$3933; H$\epsilon$\,+Ca\,{\sc ii} H $\lambda$3970; Mn\,{\sc i}
  $\lambda$4030; Fe\,{\sc i} $\lambda\lambda$4046,4271; \siii$\lambda\lambda$4128-30; \tiii+\feii double blends $\lambda\lambda$4172-8,4395-4400,4417,4444; \cai $\lambda$4227; and \mgii$\lambda$4481. Two Balmer series lines: H$\delta$ $\lambda$4101 and H$\gamma$ $\lambda$4340 are also identified.}
\label{fig:19v21}
\end{figure}

\newpage

\begin{figure}
\includegraphics[width=.9\textwidth]{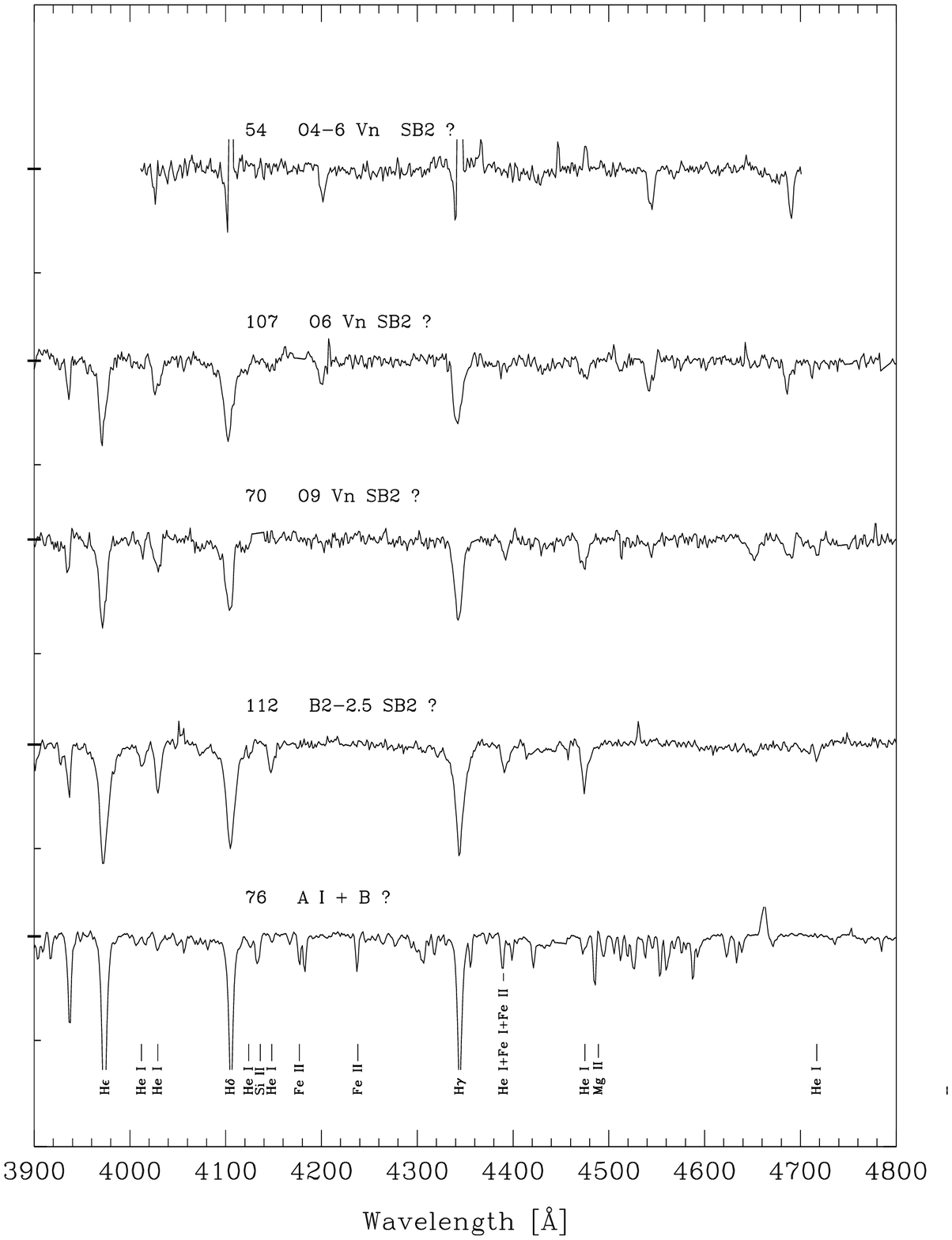}
\caption{\footnotesize Spectral classification for those spectra which show features indicating possible spectroscopic
  binary origin (See Section 3.2). Lines identified as reference are \hei $\lambda\lambda$4009,4026,4121,4144,4471,4713; \siii blend $\lambda\lambda$4128-30; \feii blend $\lambda\lambda$4172-78,4233; \mgii $\lambda$4481; and \hei $\lambda$4387 dominated by the \fei+\feii blend $\lambda\lambda$4383-85 in star 76. Three Balmer series lines: H$\epsilon$ $\lambda$3970, H$\delta$ $\lambda$4101, and H$\gamma$ $\lambda$4340 are also identified. }
\label{fig:20v21}
\end{figure}
\newpage


\begin{figure}
\includegraphics[angle=270,width=\textwidth]{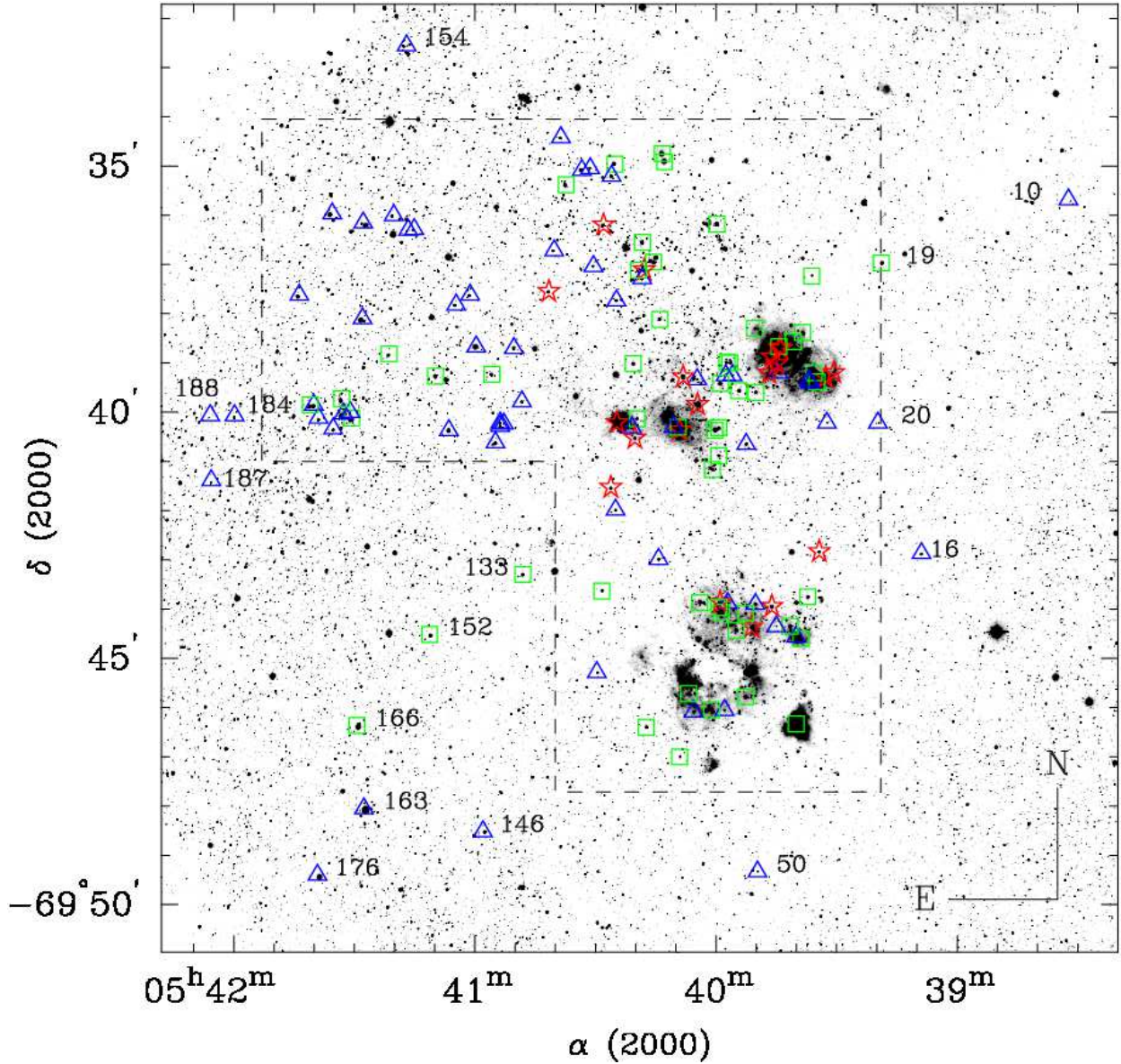}
\caption{ Spatial distribution of the stars on the sky distinguished by their spectral types, as an approximate indication of their evolutionary stages. The symbols are as follows: {\it red stars}, O3-O5~V-I and O~Vz; {\it green squares}, O6-O9~V-I and B0~I; {\it blue triangles}, B0-B2~V-III and B1-B8~I. The dashed lines contour the regions that are zoomed in Figures  \ref{fig:distributionN159}, \ref{fig:distributionN160}, and \ref{fig:distributionN160E}. Stars outside the contours are labelled with their ID numbers from Table~1. Stars inside the contours are identified in the corresponding figures.}

\label{fig:distribution}
\end{figure}

\begin{figure}
\includegraphics[angle=270,width=\textwidth]{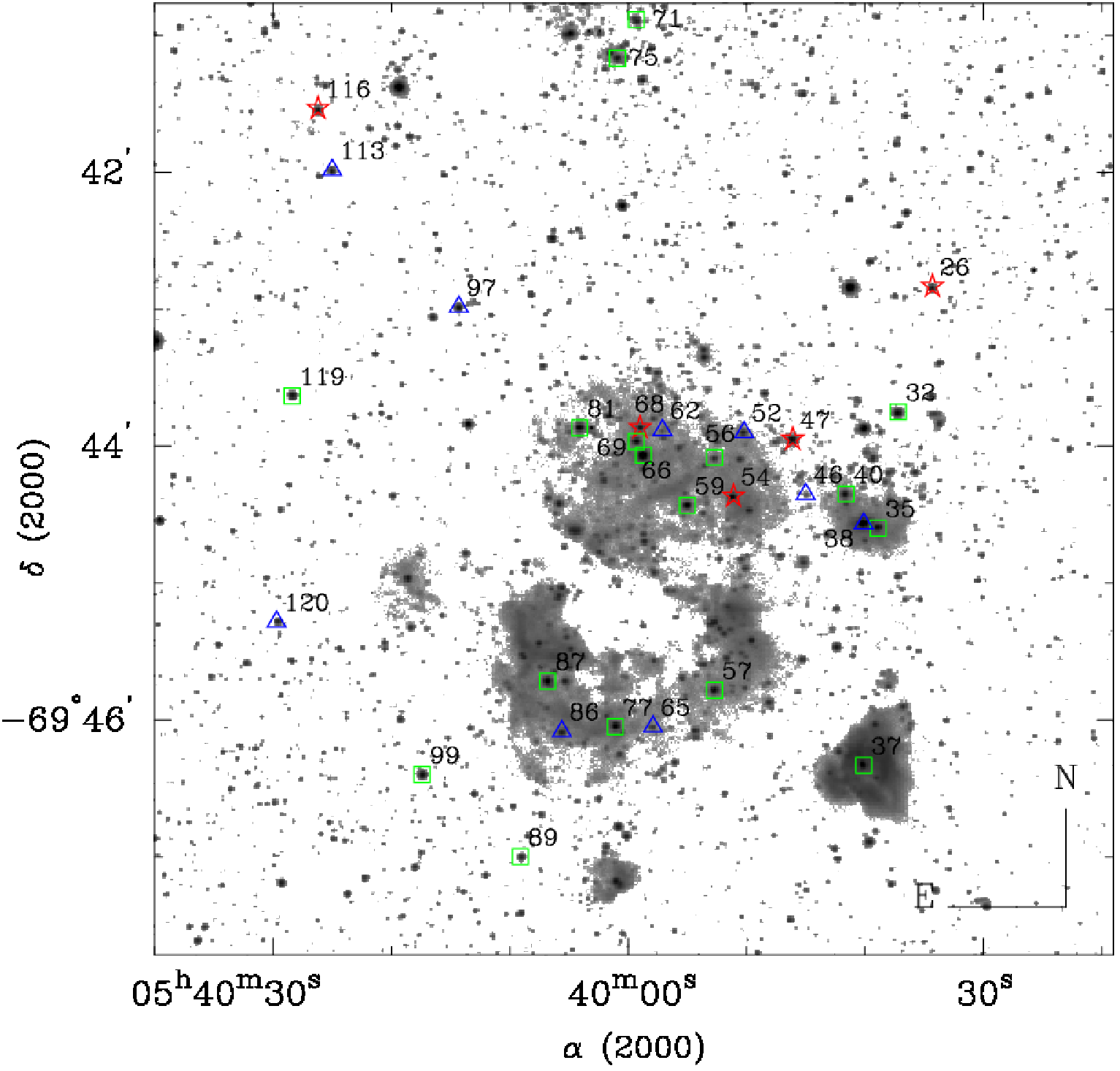}
\caption{ Spatial distribution of stars.
N159 zoomed; stars are labelled with their ID numbers from Table~1. Symbols
denote O3-O5 V-I and O Vz, {\it red stars}; O6-O9 V-I and B0 I, {\it green squares}; B0-B2 V-III and B1-B8 I, {\it blue triangles}.}
\label{fig:distributionN159}
\end{figure}

\begin{figure}
\includegraphics[angle=270,width=\textwidth]{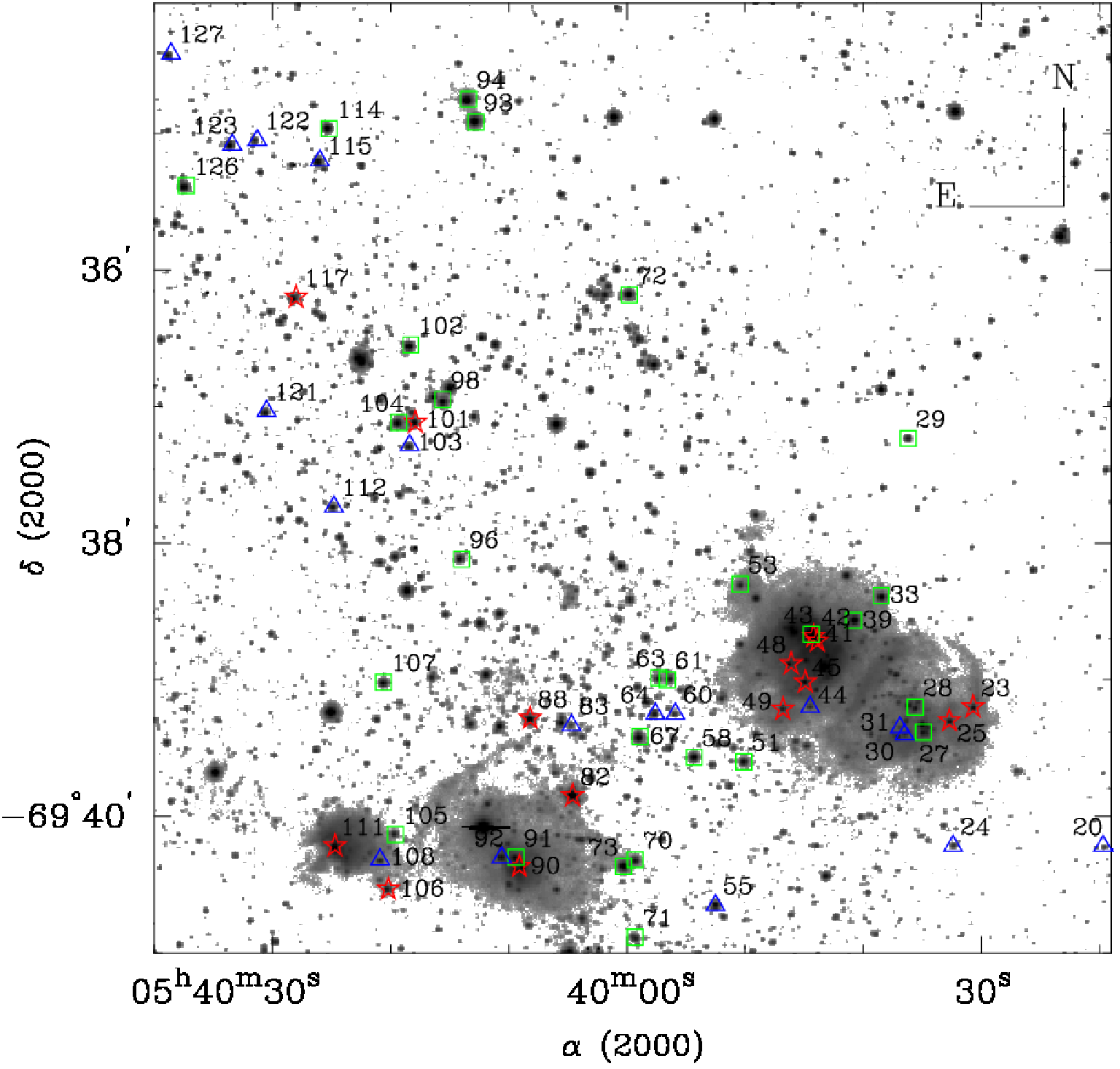}
\caption{ Spatial distribution of stars.
N160 zoomed; stars are labelled with their ID numbers from Table~1. Symbols
denote O3-O5 V-I and O Vz, {\it red stars}; O6-O9 V-I and B0 I, {\it green squares}; B0-B2 V-III and B1-B8 I, {\it blue triangles}.}
\label{fig:distributionN160}
\end{figure}

\begin{figure}
\includegraphics[angle=270,width=\textwidth]{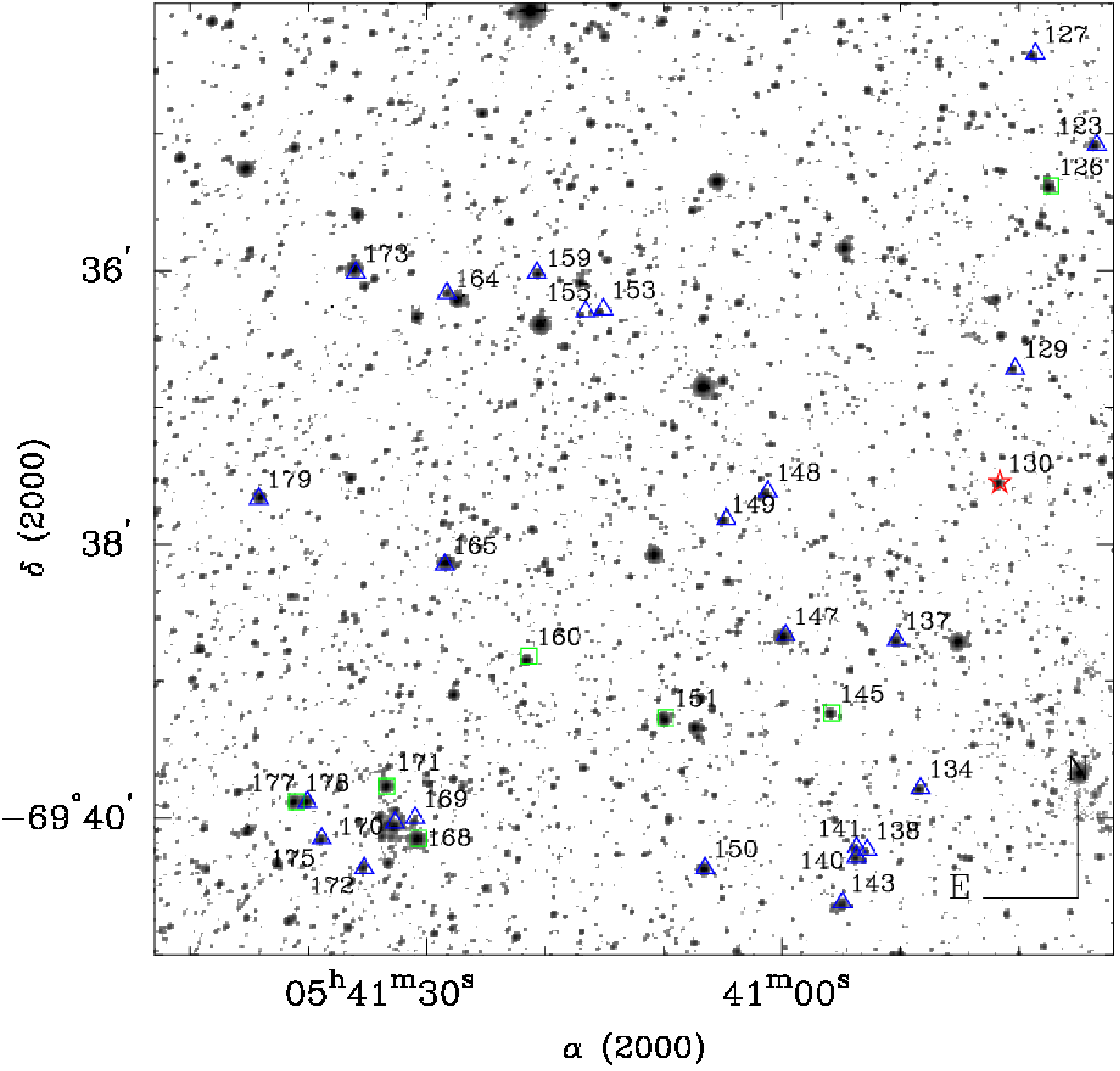}
\caption{ Spatial distribution of stars.
N160 East zoomed; stars are labelled with their ID numbers from Table~1.
Symbols denote O3-O5 V-I and O Vz, {\it red stars}; O6-O9 V-I and B0 I, {\it green squares};
B0-B2 V-III and B1-B8 I, {\it blue triangles}.}
\label{fig:distributionN160E}
\end{figure}

\bibliographystyle{apj}
\bibliography{biblio}

\begin{thebibliography}{43}
\expandafter\ifx\csname natexlab\endcsname\relax\def\natexlab#1{#1}\fi

\bibitem[{{Ardeberg} {et~al.}(1972){Ardeberg}, {Brunet}, {Maurice}, \&
  {Prevot}}]{1972A&AS....6..249A}
{Ardeberg}, A., {Brunet}, J.~P., {Maurice}, E., \& {Prevot}, L. 1972, \aaps, 6,
  249

\bibitem[{{Bianchi} \& {Pakull}(1985)}]{1985A&A...146..242B}
{Bianchi}, L. \& {Pakull}, M. 1985, \aap, 146, 242

\bibitem[{{Bolatto} {et~al.}(2000){Bolatto}, {Jackson}, {Israel}, {Zhang}, \&
  {Kim}}]{2000ApJ...545..234B}
{Bolatto}, A.~D., {Jackson}, J.~M., {Israel}, F.~P., {Zhang}, X., \& {Kim}, S.
  2000, \apj, 545, 234

\bibitem[{{Bonnarel} {et~al.}(2000){Bonnarel}, {Fernique}, {Bienaym{\'e}},
  {Egret}, {Genova}, {Louys}, {Ochsenbein}, {Wenger}, \&
  {Bartlett}}]{2000A&AS..143...33B}
{Bonnarel}, F., {Fernique}, P., {Bienaym{\'e}}, O., {Egret}, D., {Genova}, F.,
  {Louys}, M., {Ochsenbein}, F., {Wenger}, M., \& {Bartlett}, J.~G. 2000,
  \aaps, 143, 33

\bibitem[{{Bosch} {et~al.}(1999){Bosch}, {Terlevich}, {Melnick}, \&
  {Selman}}]{1999A&AS..137...21B}
{Bosch}, G., {Terlevich}, R., {Melnick}, J., \& {Selman}, F. 1999, \aaps, 137,
  21

\bibitem[{{Brunet} {et~al.}(1975){Brunet}, {Imbert}, {Martin}, {Mianes},
  {Pr{\'e}vot}, {Rebeirot}, \& {Rousseau}}]{1975A&AS...21..109B}
{Brunet}, J.~P., {Imbert}, M., {Martin}, N., {Mianes}, P., {Pr{\'e}vot}, L.,
  {Rebeirot}, E., \& {Rousseau}, J. 1975, \aaps, 21, 109

\bibitem[{{Cannon} \& {Pickering}(1918)}]{1918AnHar..92....1C}
{Cannon}, A.~J. \& {Pickering}, E.~C. 1918, Annals of Harvard College
  Observatory, 92, 1

\bibitem[{{Caswell} \& {Haynes}(1981)}]{1981MNRAS.194P..33C}
{Caswell}, J.~L. \& {Haynes}, R.~F. 1981, \mnras, 194, 33P

\bibitem[{{Conti} \& {Fitzpatrick}(1991)}]{1991ApJ...373..100C}
{Conti}, P.~S. \& {Fitzpatrick}, E.~L. 1991, \apj, 373, 100

\bibitem[{{Degioia-Eastwood} {et~al.}(1993){Degioia-Eastwood}, {Meyers}, \&
  {Jones}}]{1993AJ....106.1005D}
{Degioia-Eastwood}, K., {Meyers}, R.~P., \& {Jones}, D.~P. 1993, \aj, 106, 1005

\bibitem[{{Deharveng} \& {Caplan}(1992)}]{1992A&A...259..480D}
{Deharveng}, L. \& {Caplan}, J. 1992, \aap, 259, 480

\bibitem[{{Deharveng} {et~al.}(1992){Deharveng}, {Caplan}, \&
  {Lombard}}]{1992A&AS...94..359D}
{Deharveng}, L., {Caplan}, J., \& {Lombard}, J. 1992, \aaps, 94, 359

\bibitem[{{Feast} {et~al.}(1960){Feast}, {Thackeray}, \&
  {Wesselink}}]{1960MNRAS.121..337F}
{Feast}, M.~W., {Thackeray}, A.~D., \& {Wesselink}, A.~J. 1960, \mnras, 121,
  337

\bibitem[{{Fitzpatrick}(1991)}]{1991PASP..103.1123F}
{Fitzpatrick}, E.~L. 1991, \pasp, 103, 1123

\bibitem[{{Gatley} {et~al.}(1981){Gatley}, {Becklin}, {Hyland}, \&
  {Jones}}]{1981MNRAS.197P..17G}
{Gatley}, I., {Becklin}, E.~E., {Hyland}, A.~R., \& {Jones}, T.~J. 1981,
  \mnras, 197, 17P

\bibitem[{{Henize}(1956)}]{1956ApJS....2..315H}
{Henize}, K.~G. 1956, \apjs, 2, 315

\bibitem[{{Heydari-Malayeri} {et~al.}(2002){Heydari-Malayeri}, {Charmandaris},
  {Deharveng}, {Meynadier}, {Rosa}, {Schaerer}, \&
  {Zinnecker}}]{2002A&A...381..941H}
{Heydari-Malayeri}, M., {Charmandaris}, V., {Deharveng}, L., {Meynadier}, F.,
  {Rosa}, M.~R., {Schaerer}, D., \& {Zinnecker}, H. 2002, \aap, 381, 941

\bibitem[{{Heydari-Malayeri} \& {Testor}(1982)}]{1982A&A...111L..11H}
{Heydari-Malayeri}, M. \& {Testor}, G. 1982, \aap, 111, L11

\bibitem[{{Hutchings} {et~al.}(1983){Hutchings}, {Crampton}, \&
  {Cowley}}]{1983ApJ...275L..43H}
{Hutchings}, J.~B., {Crampton}, D., \& {Cowley}, A.~P. 1983, \apj, 275, L43

\bibitem[{{Jones} {et~al.}(2005){Jones}, {Woodward}, {Boyer}, {Gehrz}, \&
  {Polomski}}]{2005ApJ...620..731J}
{Jones}, T.~J., {Woodward}, C.~E., {Boyer}, M.~L., {Gehrz}, R.~D., \&
  {Polomski}, E. 2005, \apj, 620, 731

\bibitem[{{Liu} {et~al.}(2000){Liu}, {van Paradijs}, \& {van den
  Heuvel}}]{2000A&AS..147...25L}
{Liu}, Q.~Z., {van Paradijs}, J., \& {van den Heuvel}, E.~P.~J. 2000, \aaps,
  147, 25

\bibitem[{{Massey} {et~al.}(1995){Massey}, {Lang}, {Degioia-Eastwood}, \&
  {Garmany}}]{1995ApJ...438..188M}
{Massey}, P., {Lang}, C.~C., {Degioia-Eastwood}, K., \& {Garmany}, C.~D. 1995,
  \apj, 438, 188

\bibitem[{{Meynadier} {et~al.}(2004){Meynadier}, {Heydari-Malayeri},
  {Deharveng}, {Charmandaris}, {Le Bertre}, {Rosa}, {Schaerer}, \&
  {Zinnecker}}]{2004A&A...422..129M}
{Meynadier}, F., {Heydari-Malayeri}, M., {Deharveng}, L., {Charmandaris}, V.,
  {Le Bertre}, T., {Rosa}, M.~R., {Schaerer}, D., \& {Zinnecker}, H. 2004,
  \aap, 422, 129

\bibitem[{{Nakajima} {et~al.}(2005){Nakajima}, {Kato}, {Nagata}, {Tamura},
  {Sato}, {Sugitani}, {Nagashima}, {Nagayama}, {Iwata}, {Ita}, {Tanabe},
  {Kurita}, {Nakaya}, \& {Baba}}]{2005AJ....129..776N}
{Nakajima}, Y., {Kato}, D., {Nagata}, T., {Tamura}, M., {Sato}, S., {Sugitani},
  K., {Nagashima}, C., {Nagayama}, T., {Iwata}, I., {Ita}, Y., {Tanabe}, T.,
  {Kurita}, M., {Nakaya}, H., \& {Baba}, D. 2005, \aj, 129, 776

\bibitem[{{Pakull}(1984)}]{1984IAUS..108..317P}
{Pakull}, M.~W. 1984, in IAU Symposium, Vol. 108, Structure and Evolution of
  the Magellanic Clouds, ed. S.~{van den Bergh} \& K.~S.~D. {Boer}, 317--+

\bibitem[{{Ramsey} {et~al.}(2006){Ramsey}, {Williams}, {Gruendl}, {Chen},
  {Chu}, \& {Wang}}]{2006ApJ...641..241R}
{Ramsey}, C.~J., {Williams}, R.~M., {Gruendl}, R.~A., {Chen}, C.-H.~R., {Chu},
  Y.-H., \& {Wang}, Q.~D. 2006, \apj, 641, 241

\bibitem[{{Rousseau} {et~al.}(1978){Rousseau}, {Martin}, {Pr{\'e}vot},
  {Rebeirot}, {Robin}, \& {Brunet}}]{1978A&AS...31..243R}
{Rousseau}, J., {Martin}, N., {Pr{\'e}vot}, L., {Rebeirot}, E., {Robin}, A., \&
  {Brunet}, J.~P. 1978, \aaps, 31, 243

\bibitem[{{Sanduleak}(1970)}]{1970CoTol..89.....S}
{Sanduleak}, N. 1970, Contributions from the Cerro Tololo Inter-American
  Observatory, 89

\bibitem[{{Selman} {et~al.}(1999){Selman}, {Melnick}, {Bosch}, \&
  {Terlevich}}]{1999A&A...347..532S}
{Selman}, F., {Melnick}, J., {Bosch}, G., \& {Terlevich}, R. 1999, \aap, 347,
  532

\bibitem[{{Stock} {et~al.}(1976){Stock}, {Osborn}, \&
  {Iba\~nez}}]{1976A&AS...24...35S}
{Stock}, J., {Osborn}, W., \& {Iba\~nez}, M. 1976, \aaps, 24, 35

\bibitem[{{Testor} \& {Niemela}(1998)}]{1998A&AS..130..527T}
{Testor}, G. \& {Niemela}, V. 1998, \aaps, 130, 527

\bibitem[{{Walborn}(1971{\natexlab{a}})}]{1971ApJ...164L..67W}
{Walborn}, N.~R. 1971{\natexlab{a}}, \apj, 164, L67

\bibitem[{{Walborn}(1971{\natexlab{b}})}]{1971ApJS...23..257W}
---. 1971{\natexlab{b}}, \apjs, 23, 257

\bibitem[{{Walborn}(1973)}]{1973AJ.....78.1067W}
---. 1973, \aj, 78, 1067

\bibitem[{{Walborn}(1976)}]{1976ApJ...205..419W}
---. 1976, \apj, 205, 419

\bibitem[{{Walborn}(1977)}]{1977ApJ...215...53W}
---. 1977, \apj, 215, 53

\bibitem[{{Walborn} \& {Blades}(1997)}]{1997ApJS..112..457W}
{Walborn}, N.~R. \& {Blades}, J.~C. 1997, \apjs, 112, 457

\bibitem[{{Walborn} \& {Fitzpatrick}(1990)}]{1990PASP..102..379W}
{Walborn}, N.~R. \& {Fitzpatrick}, E.~L. 1990, \pasp, 102, 379

\bibitem[{{Walborn} \& {Fitzpatrick}(2000)}]{2000PASP..112...50W}
---. 2000, \pasp, 112, 50

\bibitem[{{Walborn} {et~al.}(2002){Walborn}, {Howarth}, {Lennon}, {Massey},
  {Oey}, {Moffat}, {Skalkowski}, {Morrell}, {Drissen}, \&
  {Parker}}]{2002AJ....123.2754W}
{Walborn}, N.~R., {Howarth}, I.~D., {Lennon}, D.~J., {Massey}, P., {Oey},
  M.~S., {Moffat}, A.~F.~J., {Skalkowski}, G., {Morrell}, N.~I., {Drissen}, L.,
  \& {Parker}, J.~W. 2002, \aj, 123, 2754

\bibitem[{{Walborn} {et~al.}(2000){Walborn}, {Lennon}, {Heap}, {Lindler},
  {Smith}, {Evans}, \& {Parker}}]{2000PASP..112.1243W}
{Walborn}, N.~R., {Lennon}, D.~J., {Heap}, S.~R., {Lindler}, D.~J., {Smith},
  L.~J., {Evans}, C.~J., \& {Parker}, J.~W. 2000, \pasp, 112, 1243

\bibitem[{{Yamaguchi} {et~al.}(2001){Yamaguchi}, {Mizuno}, {Mizuno}, {Rubio},
  {Abe}, {Saito}, {Moriguchi}, {Matsunaga}, {Onishi}, {Yonekura}, \&
  {Fukui}}]{2001PASJ...53..985Y}
{Yamaguchi}, R., {Mizuno}, N., {Mizuno}, A., {Rubio}, M., {Abe}, R., {Saito},
  H., {Moriguchi}, Y., {Matsunaga}, K., {Onishi}, T., {Yonekura}, Y., \&
  {Fukui}, Y. 2001, \pasj, 53, 985

\bibitem[{{Zaritsky} {et~al.}(2004){Zaritsky}, {Harris}, {Thompson}, \&
  {Grebel}}]{2004AJ....128.1606Z}
{Zaritsky}, D., {Harris}, J., {Thompson}, I.~B., \& {Grebel}, E.~K. 2004, \aj,
  128, 1606

\end{thebibliography}

\newpage

\label{lastpage}
\end{document}